\documentclass[11pt]{article}

\usepackage{amsfonts, caption, subcaption}
\usepackage{amsmath, amsthm, mathrsfs}
\usepackage{graphicx}
\usepackage{hyperref}
\usepackage{amssymb}
\usepackage{amsopn}
\usepackage{comment}
\usepackage{color}
\usepackage{tikz}
\usetikzlibrary{positioning,chains,fit,shapes,calc}

\usepackage[framemethod=tikz]{mdframed}
\def\ig{{\mathrm{IG}}}

\usepackage{tkz-graph}
\usetikzlibrary{arrows}

\usepackage[a4paper]{geometry}
\newtheorem{proposition}{Proposition}
\newtheorem{conjecture}{Conjecture}

\newtheorem{definition}{Definition}
\newtheorem{example}{Example}
\newtheorem{problem}{Problem}

\newtheorem{remark}{Remark}

\newcommand{\Po}{\mathsf{Poisson}}

\date{}

\begin{document}
\title{Information Theory of Molecular Communication: Directions and Challenges}
\author{Amin Gohari, Mahtab Mirmohseni,  Masoumeh Nasiri-Kenari
\\
Department of Electrical Engineering, Sharif University of Technology}
\maketitle

\begin{abstract}
 Molecular Communication (MC) is a communication strategy that uses molecules as carriers of information, and is widely used by biological cells. As an interdisciplinary topic, it has been studied by biologists, communication theorists and a growing number of information theorists. This paper aims to specifically bring MC to the attention of information theorists. To do this, we first highlight the unique mathematical challenges of studying the capacity of molecular channels. Addressing these problems require use of known, or development of new mathematical tools. Toward this goal,  we review a subjective selection of the existing literature on information theoretic aspect of molecular communication. The emphasis here is on the mathematical techniques used, rather than on the setup or modeling of a specific paper. Finally, as an example, we propose a concrete information theoretic problem that was motivated by our study of molecular communication.
\end{abstract}

\section{Introduction}
Deployment of nanodevices has profound technological implications, and opens up unique opportunities in a wide range of applications, particularly in medicine, for disease detection, control and treatment. Each nanodevice alone has a very limited operational capability. To increase their capability for complicated tasks, as required in many nano- and bio-technology applications, it is essential to envision nanonetworks and study nanoscale communication. The size and power consumption of transceivers make the electromagnetic wave based communication rather unsuitable for interconnecting nanomachines. This motivates the use of Molecular Communication (MC) as a promising communication mechanism (e.g., see \cite{Akyldiz1} or \cite{Nakano11}). Furthermore, being the prevalent communication mechanism in nature among living organisms, MC nanonetwork can use the existing micro-organisms, such as bacteria and cells, as part of network and is more compatible with the human body, a feature which is necessary for biomedical applications. We should note that even though MC was initially envisioned for nanoscale communications, there are also some potential macroscale applications for it, \emph{e.g.}, underwater communication where electromagnetic waves cannot be efficiently employed over a long distance since they experience very high attenuations  \cite{NarimanSurvey}.

MC is defined as a communication strategy that uses molecules as information carrier instead of electromagnetic waves. Information can be coded in the type, concentration or release time of molecules that are spread in the medium. As with any other practical communication medium, uncertainty, imperfection and noise exist in MC, fundamentally limiting the system performance. The existing literature on MC provides various mathematical models of a molecular communication system, each of which is, in principle, amenable to capacity calculation. Furthermore, just like classical communication, the optimal use of MC for the purposes of coordination, function computation, or control can be studied, using information theoretic tools. The idea of determining ultimate achievable limits is a helpful notion and provides an opportunity for information theorists to collaborate in the development of the theory of MC. Furthermore, MC can inspire new interesting problems for information theorists of mathematical orientation to look at.

Nanocells and nanodevices can only perform \underline{\emph{simple}} operations due to their small physical scale and limitation of resources. In his 2002 Shannon Lecture on ``Living Information Theory", Toby Berger points to the fact that living systems employ simple structures in encoding and decoding information, and have ``little if any need for the elegant block and convolutional coding theorems and techniques of information theory." Berger argues (mainly in the context of neural networks) that this is because communication medium has adapted itself to the data sources in the evolutionary process. Therefore, the optimality of encoder and decoders with simple structures is due to the fact that the channel and the data are matched. Uncoded transmission, in particular, is an appealing strategy for biological applications, and is shown to be provably optimal in some settings \cite{Gastpar}. But simplicity may find other justifications besides adaptation in the evolutionary process. Fixing the uncoded transmission, it is shown in \cite{Ourwork1} that a certain memory-limited simple decoder is performing close to the optimal decoder. Conversely, in \cite{Ourwork2}, we fix a simple decoder and show that a certain memory-limited simple encoder is near optimal. These results seem to suggest that even though the optimal transmitter and Maximum Likelihood (ML) decoder may have complicated descriptions, the nature of the molecular channel is such that \emph{simplicity propagates}:  if some components of a MC system are forced to be structurally simple (due to their physical limitation), then using complicated coding strategies at other components comes at a negligible benefit.

Differences between classical and molecular communication open up the possibility of defining new problems for information theorists. Some of these differences are as follows:

\textbf{Complexity:} MC may be used for both microscale and macroscale applications. In the case of microscale applications, complexity is a more serious issue compared to classical communication due to the small scale of nanodevices. Nanodevices are simple and resource limited devices. An important question is how to find a proper theoretical framework  for studying the limitation of computational resources in the context of MC. So far, the molecular communication literature has treated complexity in a loose manner. Simplicity is generally invoked to justify certain restrictions of molecular encoders and decoders to a class of intuitive and easy-to-analyze functions. Unfortunately, classical information theory does not accommodate for a quantitative restriction on the degree of simplicity (limitations of computational capacity or memory) of the encoder and decoder. Coding theory aims to find practical capacity achieving codes with affordable encoder and decoder complexity. Furthermore, finite blocklength and one shot results in information theory relate to complexity. Nonetheless, proving fundamental lower bounds on the complexity can be a very difficult problem, and computational formulations such as the ones given in \cite{Yao82a, HILL99}  are too formal and abstract. The progress has been mainly within the context of specific circuit models, \emph{e.g.}, authors in \cite{ElGamalGreene} consider the VLSI model to estimate the complexity of the implementation of an error correcting code (see also \cite{Grover1, Grover2} for further computational results based on the VLSI circuit model). To sum this up, the development of a similar ``molecular circuit model" seems to be the most promising direction to address the computational aspects of MC.

\textbf{Nature of transmitter and receiver}: Even when there is no channel noise, the capacity of a MC system is constrained by the physics of transmitter and receiver: the transmitter's \emph{actuation} in response to excitement can be imperfect; the receiver may have a fundamental \emph{sensing noise}, which is independent of the channel noise. For instance, in ligand receptors where incoming molecules bind with receptors on the surface of the receiver, the sensing noise has a variance that is dependent on the amplitude of the signal \cite{EinSarFekISIT12}, \emph{i.e.}, the higher the amplitude of the signal, the larger the variance of its observation noise. Also, both the transmitter and receiver may be allowed to actively modify the communication medium itself by releasing chemicals in the environment.
Furthermore, similar to classical communication, the transmitter and receiver may be mobile, causing a change in the effective channel between the transmitter and the receiver. However, unlike the classical communication, the direction of mobility may itself be influenced by the concentration of molecules released in the environment by other nodes (as in chemotaxis of many cells).

\textbf{Use of multiple molecule types:} in MC, we can employ multiple molecule types for signaling. A classical analogue of this degree of freedom is frequency: channels of different molecule types can correspond to channels over different frequencies. But there are some crucial limitations to this analogy: unlike waves moving on different frequencies, molecules of different types might undergo chemical reactions with each other as they travel from the transmitter to the receiver. These reactions among different molecule types can result in a nonlinear channel \cite{FarsadGoldsmith}. Furthermore, molecules of different types might compete with each other at the receiver in terms of binding with receptors on the surface of the receiver; if a receptor bonds with one molecule type, it will be unable to bond with other molecule types for a period of time.

\textbf{Positivity of the input signal:} the linearity and time-invariance of wireless channel enables one to borrow tools from linear algebra or Fourier analysis. Macroscopic diffusion in a stable medium also results in a linear and time-invariant system. However, we cannot readily use tools from Fourier analysis: unlike electromagnetic waves whose amplitude can become negative, only a non-negative concentration of molecules can be released in the environment. The non-negativity constraint in the time domain does not have an easy equivalent in the Fourier domain. To simulate negative signals, authors in \cite{FarsadGoldsmith} suggest exploiting chemical reactions to reduce the concentration of a molecule type. Unfortunately, diffusion with chemical reactions follows a \emph{non-linear} differential equation, prohibiting the use of linear theories (see \cite{MosayebiIWCIT16} for a partial solution based on the fact that even though the concentration of each molecule type follows a non-linear differential equation, the difference of the concentrations still follows a linear differential equation under some assumptions).

\textbf{Energy limitation:} in MC, in contrast to the classical communication, some energy is required to synthesize a molecule. While increasing the concentration of released molecules increases the channel capacity, the amount of energy consumed for their synthesis and transmission in an active transport MC channel increases as well. Thus, for the energy limited MC system, by taking into account the energy required for synthesizing the molecules, there exists an optimum number of released molecules \cite{NarimanSurvey}. The classical analogue of this (for electronic circuits) is given in \cite{Grover1} where it is shown that approaching Shannon's capacity may require very large energy consumption.

\textbf{Slow propagation:} when studying MC via diffusion in an aqueous or gaseous medium, we should note that the speed of transmission is slow. The slow propagation in conjunction with possible changes in the medium has implications in terms of mathematical modeling of the problem. For instance, this can make it difficult to obtain channel state information at the encoder via a feedback link from the decoder. Next, because of the slow propagation of released molecules, self-interference among successive transmission is a challenge. As stated above, the concentration of molecules  in an environment is a non-negative quantity. This prevents the employment of some classical techniques for capacity evaluation, such as Fourier transform to convert an inter-symbol-interference (ISI) channel to a parallel memoryless channel.

\textbf{Focus of this paper:} There are many existing works that address various aspects of MC, sometimes from an information theoretic perspective. We are selective in reviewing these works. Our focus are on the works that are more theoretical and appealing to pure information theorists. For instance, there are many works that model a molecular communication system  with a memoryless channel and then evaluate the capacity of the resulting channel. While valuable because of their modeling aspects, their analysis may not be exciting to information theorists. We are more interested in works that do not just borrow and apply tools from information theory, but can rather attract information theorists and help form a dialogue between molecular communication and information theory. There is one more caveat: we do not review a few number of works (mostly published in biological journals) that explain evolution of biological structures by arguing that a certain information theoretic criterion is optimized.\footnote{For instance, many cells move according to spatial differences in the concentration of certain chemicals (chemotaxis). In \cite{Andrews}, an information theoretic criterion is proposed to model how cells find their migration direction from the imperfect information they obtain through chemical receptors on their surface.}

Genomics and molecular biology is listed as one of the future research directions in information theory by a number of information theorists \cite{FutureIT}, even though molecular communication is not specifically mentioned. Nonetheless, molecular communication continues to attract the attention of more information theorists (as evidenced by sessions dedicated to it in the ISIT conferences), and our hope is that this paper encourages more to join.

This paper is organized as follows: we begin by reviewing transmitter, channel and receiver models for MC  in Section~\ref{sec:model}. Depending on the choice of the transmitter and receiver model, a number of end-to-end models are given in Section~\ref{sec:end-to-end-rev}.
Next, a section is devoted to each of the end-to-end models: in Section~\ref{sec:concen} and Section~\ref{sec:rel_time} we review capacity results for a transmitter that puts information on the concentration, and on the release time of molecules, respectively. We also review the results on the capacity of the ligand-receptor in Section~\ref{sec:ligand}. Next, we turn to a multi-user setting in  Section~\ref{sec:cascade}. Molecular channels have memory, and capacity of network of channels with memory is of relevance to MC. In Section~\ref{sec:cascade}, we pose and discuss (in detail) the problem of finding the capacity of a cascade of channels with memory. Finally, we present some concluding remarks in Section~\ref{sec:concl}. Some of the proofs are moved to appendices.

\textbf{Notation:} Throughout, we use capital letters to denote random variables and small letters to denote their values. The set $\{1,2,\ldots, n\}$ is shown by $[1:n]$. The sequence $(x_1, x_2, \ldots, x_n)$ is shown by $x^n$. The input to the channel is generally denoted by rv $X$, and the output is denoted by either $Y$ or $Z$.

\section{System Model}\label{sec:model}
\begin{figure}
\begin{center}
\includegraphics[width=1\textwidth]{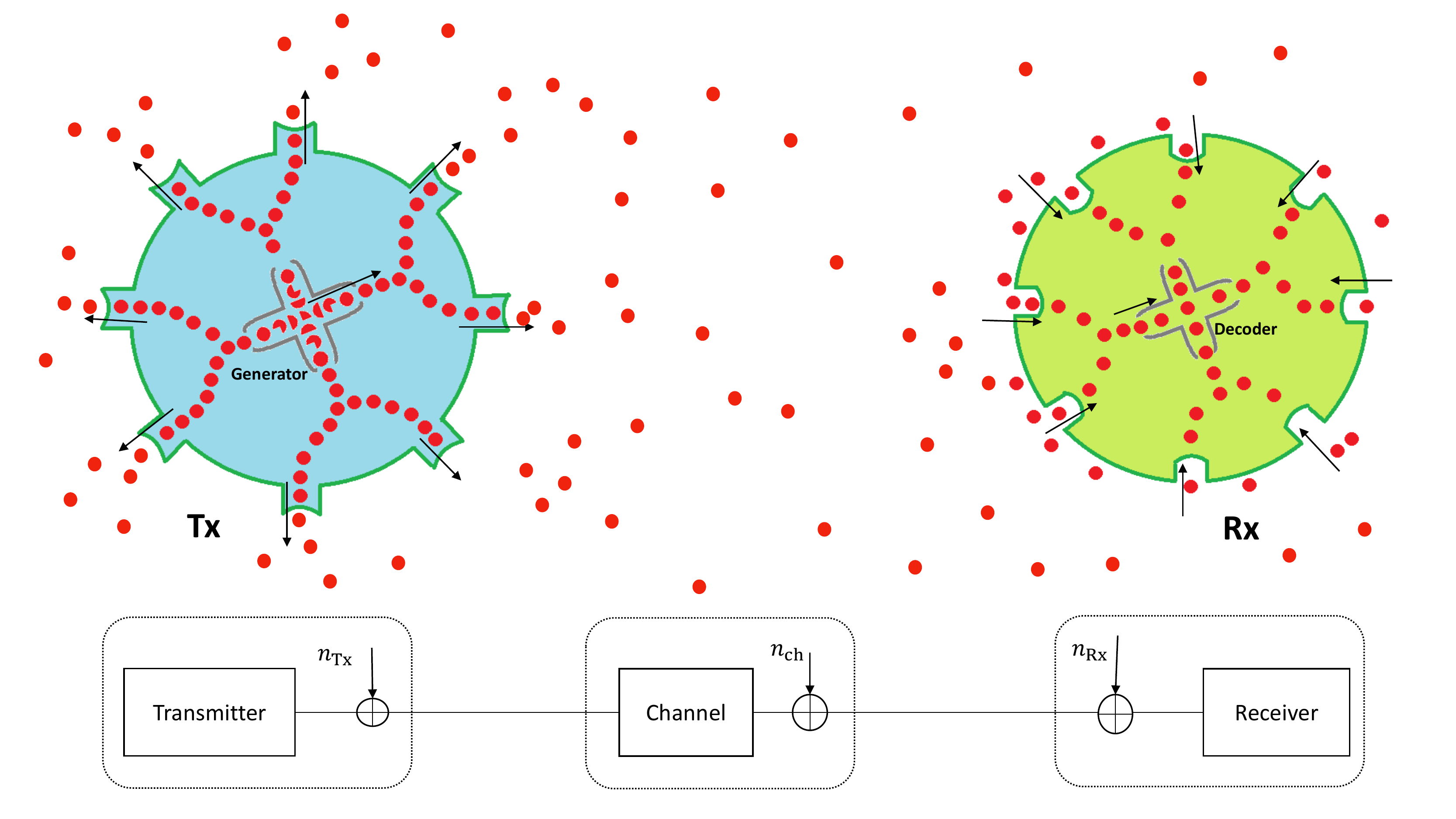}
\end{center}
\caption{Schematic illustration of a point to point molecular communication system consisting of three main componets: transmitter, channel and receiver. Each of these components has its own inherent imperfection and noise.}\label{system_model}
\end{figure}

A point to point engineered communication system consists of a transmitter, a channel, and a receiver, as shown in Fig.~\ref{system_model}. We employ this structure in our review of a molecular communication system. A molecular transmitter is a biological or engineered cell whose actions influence the density of molecules in the environment (generally by emitting molecules in the environment).\footnote{
The transmitter may change the communication medium instead of emitting molecules. See Secion \ref{sec:concl} for a discussion.}
 A molecular channel refers to a physical medium in which molecules propagate. A molecular receiver (or sensor) is a biological or engineered cell that is influenced by the density of molecules in the environment at its vicinity.

In the following three subsections, we discuss the molecular transmitter, channel and the receiver, separately. Besides this, observe that the physical structure of the transmitter or receiver imposes some limitations and imperfections on the transmission and reception processes. Therefore, it is also possible to lump together the imperfections of the transmitter, channel, and receiver and define an end-to-end channel model. We discuss the end-to-end models in the final subsection. Despite the fact that there are established models for the diffusion process, modeling the uncertainty originated from the  transmitter and the receiver is still an open area for research.

\subsection{Transmitter}
A transmitter may be modeled as a point source of molecules, located at the origin. The transmitter can control the concentration, the type, and the release time of molecules at its location.
As an example, information can be encoded in the concentration by releasing nothing for information bit $0$, and releasing a given concentration of molecules for bit $1$ (on-off keying); information can be encoded in the type by releasing molecules of type $A$ for bit $0$, and molecules of type $B$ for bit $1$. Finally, information can be encoded in time by adjusting the release time of consecutive molecules based on the input bit.

Released molecules diffuse in the environment. The number of released molecules depends on the distance between the transmitter and receiver. If the distance between the transmitter and receiver is very small, individual molecules may be transmitted one by one. If the distance is large, molecules get diluted in the environment before reaching the receiver, and the transmitter may need to send considerably more molecules to ensure a viable communication line to the receiver.

\emph{Transmitter imperfection:} In practice, the transmitter cannot perfectly control the number or the release time of the molecules. Furthermore, the molecule generation process of the transmitter can impose its own inherent constraints on the transmitter. To model these imperfections, one has to know the exact physical implementation of the transmitter. For instance, a physical description of a transmitter is detailed in \cite[Section III.A]{PierobonAkyildiz2011}. In \cite{Chou2015}, different chemical reactions are considered for the emission of different symbols.  In \cite{Ourwork1}, the transmitter is assumed to have a reservoir of molecules, with an outlet whose size is controlled by the transmitter. In other words, the transmitter may not control the exact number of molecules that exit the reservoir, but only the size of its outlet. Given the large number of molecules in the reservoir and a small probability of each exiting through the reservoir, the number of molecules exiting the outlet can be assumed to follow a Poisson distribution. Therefore, in this model, the number of released molecules from the transmitter is a Poisson random variable, where its average or rate is determined by the transmitter \cite{Ourwork1}. The Poisson model also arises in the following contexts:
\begin{itemize}
\item  Poisson distribution models the number of escaped particles from a bounded domain when gates on the boundary of the domain open and close randomly, \emph{e.g.} as in the escape of diffusing proteins from a corral in the plasma membrane \cite[Sec. 3.6]{Biobook}. Similarly, it arises in \cite{ArjAhmSchKen16}, wherein molecule release based on ion channels across the cell membrane is considered; the opening and closing of these channels are controlled by a gating parameter.
\item  It also arises when transmitter uses a colony of bacteria for molecule release. Each receptor on a bacterium  releases a molecule with a probability that depends on the amount of provocation  by the transmitter. If $p$ is the release probability and $N$ is the total number of colony receptors, the number of released molecules follows a binomial distribution with parameters $N$ and $p$. This can be approximated by a Poisson distribution if $N$ is large and $p$ is small.
\end{itemize}

\emph{Transmitter models:} In this paper we focus on the following transmitter models:
\begin{itemize}
\item \emph{Timing transmitter}: transmitter releases individual molecules one by one, at specified time instances.
\item \emph{Exact concentration transmitter:} A time slotted transmission strategy is employed: time is divided into intervals of length $T_s$, with transmission occurring at the beginning of each time slot.\footnote{While this is a common assumption, it is possible to study communication rates via a continuous-time/amplitude model for the channel \cite{Mahdavifar}. This is the only paper that we are aware of that studies mutual information terms between input and output in continuous time domain.}  The transmitter releases exactly $X_i$ molecules (or $X_i$ moles of molecule if the communication is very long range) at the beginning of the $i$-th time slot, \emph{i.e.,} at time $iT_s$.
\item \emph{Poisson concentration transmitter:} Again a time slotted transmission strategy is employed.  The number of released molecules from the transmitter  at the beginning of the $i$-th time slot  is a Poisson random variable with mean $X_i$.
\end{itemize}

\subsection{Channel}
To obtain a statistical model for a molecular channel, one has to specify the physical mechanism of molecule transport between the transmitter and the receiver. The physical motion of molecules towards the receiver may be walk-based, flow-based, or diffusion-based \cite{NarimanSurvey}. In \textbf{walk-based mechanisms}, information molecules are encapsulated into a cargo, which then by a motor protein, such as dynein and kinesin, are pushed toward the destination through a pre-defined path, like microtubule tracks. It is an active propagation and requires chemical energy (ATP). In \textbf{flow-based mechanisms}, molecule propagation is influenced by an external flow, like propagation of hormones in the blood stream. Flow is a one-way phenomenon, which makes it unsuitable for a two-way communication.
In contrast, in the \textbf{diffusion-based transport}, the molecules randomly propagate in all available directions via Brownian motion and have the most spontaneous motion. This results in a higher degree of uncertainty at the receiver, compared to the other mechanisms.
The diffusion-based mechanism is completely passive and always available without any energy cost or prior infrastructure, and is mostly suitable for highly dynamic and unpredictable environments.  It is also possible to consider diffusion-based transport in the presence of a drift, resulting in a mixture of diffusion and flow based mechanisms.
Most of the literature (as well as this paper) is focused on the diffusion-based transport mechanism.

Molecular diffusion can be studied from either \emph{microscopic} or \emph{macroscopic} points of view. In a {microscopic} point of view, the focus is on the random movement of individual molecules,  known as the \emph{Brownian motion}. In a {macroscopic} point of view, the focus is on the overall behavior of an enormous number of molecules. Even though each molecule still has a random movement, but the total average behavior is characterized by a deterministic differential equation (due to the law of large numbers phenomenon); this deterministic differential equation is called the \emph{Fick's law of diffusion}. Historically, Fick proposed his macroscopic law of diffusion in 1855, while the microscopic Brownian motion was studied later by Einstein in 1905; a full mathematical theory based on the theory of stochastic differential equations was developed only later in the 20th century.

\subsubsection{Macroscopic diffusion}
According to Fick's first law of diffusion, diffusion flux goes from the region of high concentration to the region of low concentration. Furthermore, the diffusion rate, denoted by $J$, is proportional to the concentration gradient.
For simplicity of exposition, let us consider one-dimensional diffusion \footnote{Equations describing the three-dimensional diffusion are similar to those describing the one-dimensional diffusion. Hence, to convey the intuition we start with a one-dimensional diffusion.}. Then, Fick's first law is:
\begin{align}{\displaystyle J(x,t)=-D{\frac {\partial \rho(x,t) }{\partial x}}}\label{eqnFickrev1}\end{align}
where
$J(x,t)$ and $\rho(x,t)$ are respectively the diffusion flux, and molecule concentration (in molar) at location $x$ at time $t$; $D$ is the diffusion coefficient of the environment.

Now, assuming that there are no chemical reactions, drift velocity, or injection of external molecules to the environment, we can use the mass conservation principle to conclude that
\begin{align}\frac{\partial \rho(x,t)}{\partial t} =-\frac{\partial J(x,t)}{\partial x}.\label{eqnFickrev2}\end{align}
To intuitively understand the above equation, for a $\Delta_x>0$, $J(x,t)-J(x+\Delta_x,t)$ shows the entry rate of molecules  from position $x$ minus the exit rate of molecules from position $x+\Delta_x$; if there is a mismatch between the entry and exit rates, the total number of molecules in the interval $[x, x+\Delta_x]$ changes at a rate equal to the difference of the entry and exit rates, \emph{i.e.,}  at rate $J(x,t)-J(x+\Delta_x,t)$.

Equations \eqref{eqnFickrev1} and \eqref{eqnFickrev2} give us Fick's second law of diffusion:
\begin{align}\frac{\partial \rho(x,t)}{\partial t} =D\frac{\partial^2}{\partial x^2}\rho(x,t).\label{eqnFickrev21}\end{align}
If in addition to the diffusion process, we also have a molecule production rate, the changes in molecule concentration will be both as a result of molecule production as well as diffusion,
\begin{align}\frac{\partial \rho(x,t)}{\partial t} =-\frac{\partial J(x,t)}{\partial x}+c(x,t).\label{eqnFickrev3}\end{align}
Here, for a given time $t$, $c(x,t)$ denotes the density of molecule production rate at point $x$, \emph{i.e.,} the number of molecules added to the environment in $[x, x+dx]$ between time $[t, t+dt]$ is equal to $c(x,t)dxdt$. Then, we can write Fick's second law as
$$\frac{\partial \rho(x,t)}{\partial t} =D\frac{\partial^2}{\partial x^2}\rho(x,t)+ c(x,t).$$
To solve this differential equation in an interval $\mathscr{C}=[a,b]$, it suffices to know the initial and boundary conditions: the initial density $\rho(x,0)$ at time zero for $x\in \mathscr{C}$. The boundary condition imposes some constraints on $\rho(a,t)$ and $\rho(b,t)$ as follows:
\begin{itemize}
\item Known values of $\rho(x,t)$ at $x=a,b$ for all $t>0$: this corresponds to known concentration on the boundaries. A special case of this is the zero boundary condition: $\rho(a,t)=\rho(b,t)=0$; this corresponds to an absorbing boundary, \emph{i.e.,} molecules are absorbed and removed from the environment upon hitting the boundary of $\mathscr{C}$.
\item Known values of $\frac{\partial \rho(x,t)}{\partial x}$ at $x=a,b$  for all $t>0$: this corresponds to known diffusion flux $J(x,t)$ on the boundaries. A special case of this is the zero boundary condition: $J(a,t)=J(b,t)=0$; this corresponds to a reflecting boundary, \emph{i.e.,} molecules that hit the boundary walls are reflected back into $\mathscr{C}$.
\end{itemize}
When $\mathscr{C}=(-\infty,\infty)$ is the entire real line, the boundary can be placed as we take the limits to  infinity; for instance, one can solve the differential equation assuming that $\rho(x,t)$ vanishes at infinity (as $x$ becomes large). As an example,  consider a transmitter that is located at the origin ($x=0$), and at time $t=0$ suddenly releases one unit of molecules in the environment. In this case, the molecule production rate will be equal to $c(x,t)= \delta(x=0)\delta(t=0)$. In response to this input, the output $\rho(x,t)$ from equation \eqref{eqnFickrev3} (assuming vanishing $\rho(x,t)$ at infinity) is equal to the Green's function:
\begin{align}
\rho(x, t)=\frac{\mathbf{1}[t>0]}{(4\pi Dt)^{0.5}}  e^{-\frac{x^2}{4Dt}}.\label{eqn:impulse1d}
\end{align}
Observe that for a fixed $t$, $\rho(x, t)$ as a function of $x$ is the pdf of a Gaussian distribution with variance $2Dt$. This is no coincidence, and will become more transparent once we consider the microscopic interpretation of the diffusion process.

If there is a drift, in addition to pure diffusion, influencing the molecules motion, the modified Fick's laws may be employed to analyze the molecular channel. Solving the differential equations representing Fick's laws can be very cumbersome when non-ideal assumptions are considered, \emph{e.g.}, non-homogeneous environment, bounded space, or turbulent diffusion. Thereby, obtaining explicit channel models for molecular communication becomes complicated. A more difficult condition occurs when we consider multiple diffusing molecules subject to chemical reactions. While one can still find the differential equations that describe the process, a closed form analytical solution may not exist.\footnote{Nonetheless, diffusion and reaction-diffusion differential equations have been subject to numerious studies; see for instance the first two chapters of \cite{Pao}. For instance, much is known about existence of their solutions, their global boundedness, stability and asymptotics.}

\subsubsection{Microscopic diffusion} \label{micro-section}
In his celebrated work in 1905, Einstein showed that the density function of the movement of a single particle under Brownian motion satisfies the differential equation given by Fick's second law of diffusion. There are different equivalent ways to formally define the Brownian motion, which is a continuous time, random-walk process; \emph{e.g.,} compare \cite{Feller} and \cite[p.16]{Protter}. The one-dimensional Brownian motion $B(t), t\geq 0$ can be defined as follows \cite[p.16]{Protter}: (i) future displacements of the particle are independent of past movements. In other words, for $0\leq t_1<t_2$, $B(t_2)-B(t_1)$ is independent of $\{B(t), t\in[0:t_1]\}$; (ii) increments are normally distributed, \emph{i.e.,} for $0\leq t_1<t_2$, $B(t_2)-B(t_1)$ is a Gaussian variable $\mathcal{N}(0, 2D(t_2-t_1))$. The definition of Brownian motion is self-consistent, because sum of independent normal variables is also a normal variable. Now, assume a single particle released at time zero, at the origin: $B(0)=0$. Let us denote the distribution of the location of the particle at time $t$ by $\rho(x, t)$. Then,  the particle at time $t$ follows a normal distribution with variance $2Dt$, and $\rho(x,t)$ has the same value as given in equation \eqref{eqn:impulse1d}.

Both probability density function of a single particle (microscopic), and the concentration profile of molecules (macroscopic) satisfy Fick's diffusion law, even though these two are conceptually different. For instance unlike the integral of a density function, the integral of the  concentration on the entire space (\emph{i.e.}, the total number of molecules) can be greater than one. Next, note that to solve Fick's differential equation, boundary conditions are also needed. These can be imposed in the microscopic perspective in same manner as they are imposed in the macroscopic perspective. Then, the distribution of the particle at time $t$ can be found by solving Fick's law of diffusion with proper boundary conditions.

We refer the reader to \cite[Section 2.2]{Roisin}  for an illustrative discussion of Einstein's connection between the Brownian motion random walk and diffusion. For a rigorous mathematical discussion, see \cite{Oksendal, Feller}. But to explain the connection at an intuitive level, assume that a very large number of molecules, $N$ molecules, are released at time zero at the origin. These molecules move randomly and independently of each other. Assume that each molecule falls into the interval $[x, x+\Delta(x)]$ with probability $\rho(x,t)\Delta(x)$. Then, since there are $N$ molecules and each molecule falls in $[x, x+\Delta(x)]$ independently of other molecules,  the number of molecules that  fall in  $[x, x+\Delta(x)]$ is distributed according to a binomial distribution with parameters $(N, \rho(x,t)\Delta(x))$. The expected value of this random variable is equal to $N\rho(x,t)\Delta(x)$. By the law of large numbers, for a fixed $\Delta(x)$, if we let $N$ go to infinity, the binomial distribution has a sharp concentration around this expected value. For instance, if $N$ is
one mole of molecules ($N=6.022\times 10^{23}$), then $N\rho(x,t)$ can be expressed as $\rho(x,t)$ molar. Then, the mapping $(x,t)\mapsto \rho(x,t)$ also gives us the average macroscopic concentration profile of molecules as a function of $t$ and $x$ in terms of molar. On the other hand, we know that the macroscopic concentration satisfies the Fick's laws of diffusion. As a result, $\rho(x,t)$ satisfies the Fick's laws of diffusion.

The above argument also provides a bridge between the macroscopic and microscopic perspectives: the macroscopic concentration is $\tilde{\rho}(x,t)=N\rho(x,t)$, where the microscopic probability density is $\rho(x,t)$, and the number of molecules that fall into interval $[x, x+\Delta(x)]$ is distributed according to a binomial distribution with parameters $(N, \rho(x,t)\Delta(x))$. This binomial distribution may be approximated by Gaussian or Poisson distributions \cite{ArjAhmSchKen16,Arjmandi2013,PieAky11,MahMakMou14,NoeCheSch14,YilHerTugCha14, Ghavami}. In case of
large $N\rho(x,t)\Delta(x)(1-\rho(x,t)\Delta(x))$ (even for small $\Delta(x)$), the binomial distribution can be approximated with a Normal distribution with mean $N\rho(x,t)\Delta(x)$ and variance $N\rho(x,t)\Delta(x)(1-\rho(x,t)\Delta(x))$ \cite[p.80]{RossBook}. With the assumption that $\rho(x,t)\Delta(x)$ is small, the variance $N\rho(x,t)\Delta(x)(1-\rho(x,t)\Delta(x))$ can be approximated with  $N\rho(x,t)\Delta(x)$. Then, the density of molecules, \emph{i.e.,} the number of molecules divided by $\Delta(x)$, follows a Normal distribution with mean $N\rho(x,t)=\tilde{\rho}(x,t)$ and variance $N\rho(x,t)\Delta(x)/\Delta(x)^2=\tilde{\rho}(x,t)/\Delta(x)$. To sum this up, if the macroscopic perspective on diffusion predicts a concentration $\tilde{\rho}(x,t)$ molecules per volume, the actual concentration of molecules per volume (in an interval  $\Delta(x)$) is a normal random variable whose mean is $\tilde{\rho}(x,t)$, and whose variance is $\tilde{\rho}(x,t)/\Delta(x)$. 

\begin{example}\label{example:rev1}
Fick's law of diffusion in terms of probability demonstrates how the distribution of the particle's location at time $t$, denoted by $\rho(x,t)$, changes over time. To show the use of this fact, consider a transmitter located at the origin $x=0$, and an absorbing receiver located at $x=d$, at distance $d$ from the transmitter. An absorbing receiver is particularly important in molecular communication and the result of the following derivation is used elsewhere in the paper.

Assumed that a single molecule is released at time zero, therefore the initial distribution of the particle at $t=0$ is $\rho(x,0)=\delta(x)$. The molecule disappears upon hitting the receiver. Solving for Fick's equation with the boundary conditions  $\rho(d,t)=0$ and $\lim_{x\rightarrow -\infty}\rho(x,t)=0$, we can obtain $\rho(x,t)dx$ as the probability that the particle is located in $[x, x+dx]$ at time $t$.  We refer the reader to \cite[Section 2.5]{Roisin} for the technique for solving  Fick's differential  \eqref{eqnFickrev3}  with this boundary condition. Then, the probability that the particle has not hit the receiver by time $t$ is equal to
$$\int_{x=-\infty}^{d}\rho(x,t)dx.$$
In other words, if $T$ denotes the hitting time of a released particle with the absorbing receiver, the distribution of $T$ can be  obtained as follows:
$$\mathbb{P}[T\geq t]=\int_{x=-\infty}^{d}\rho(x,t)dx.$$
In case of the diffusion equation given in \eqref{eqnFickrev21} in a one-dimensional free homogeneous medium with diffusion coefficient $D$, the first arrival time $T$ at the receiver can be explicitly calculated. First shown by Schrodinger in 1915, $T$ has a L\'evy distribution \cite{Timing4n}
\begin{equation}
	\label{eqn:Levy}
	f_T(t) = \left\{
		\begin{array}{cc}
			\sqrt{\frac{\lambda}{2 \pi t^3}}
			\exp \left( -\frac{\lambda}{2t} \right) , & t > 0; \\
			0, & t\leq 0 .
		\end{array}
	\right.
\end{equation}
where $
	\lambda={d^2}/{2D}.$
On the other hand, if the medium also has a velocity $v$ towards the receiver, the modified Fick's law can be used to show that $T$ follows an inverse Gaussian distribution, $\ig(\mu,\lambda)$ \cite{Timing3}:\footnote{A similar distribution holds for diffusion in a free three-dimensional environment \cite{Timing4n,TimingAux1}
}:
\begin{equation}
	\label{eqn:IG}
	f_T(t) = \left\{
		\begin{array}{cc}
			\sqrt{\frac{\lambda}{2 \pi t^3}}
			\exp \left( -\frac{\lambda (t - \mu)^2}{2 \mu^2 t} \right) , & t > 0; \\
			0, & t\leq 0 .
		\end{array}
	\right.
\end{equation}
where
\begin{eqnarray}
	\mu & = & \frac{d}{v} , \:\: \mathrm{and} \\
	\lambda & = & \frac{d^2}{2D} .
\end{eqnarray}
Note that L\'evy is a heavy-tailed distribution (has  infinite mean and variance), while IG has an exponentially decreasing tail.
\end{example}
\label{sec:mic}

\subsubsection{Statistical models of molecular channel}
To construct a statistical model for a molecular channel, the microscopic and macroscopic views of the diffusion should be appropriately utilized. On the transmitter side, if only a few molecules are released, the microscopic perspective is relevant. However, if a large number of molecules are released by the transmitter, the macroscopic perspective becomes relevant. On the other hand, in case of a single tiny molecular receiver, the microscopic behavior of molecules around the receiver is of importance. As a result, if a large number of molecules are released by the transmitter and the receiver is a single tiny nanomachine, we can employ the macroscopic view to compute the molecular concentration around the receiver, but then switch to the microscopic view  as discussed in Section \ref{sec:mic}, and consider the actual number of molecules around the receiver. This is done explicitly later in Section \ref{sec:end-to-end-rev}.

Finally, we comment that while in the above we restricted to one-dimensional diffusion, similar formulas hold for diffusion in three dimensions. While Fick's law have extensions for nonhomogeneous environment with general boundary conditions,  most existing works study the simple case of an infinite medium in each direction, with no barrier or obstacle except the receiver surface.

\subsection{Receiver}\label{sec:rece:rev}
There are different models of the receiver in the literature:
\begin{itemize}
\item \emph{Sampling receiver:} The simplest model for a receiver is a device that measures the macroscopic molecular concentration at a given point. The receiver's observation is assumed not to affect the diffusion of the molecules, hence it imposes no boundary conditions when solving Fick's law of diffusion. The receiver may be assumed to sample the medium at certain given time instances.
\item \emph{Transparent receiver:} In this model, the receiver is a transparent sphere of volume $V_R$ rather than a point, but still not affecting the diffusion of the molecules. Hence it imposes no boundary conditions when solving Fick's law of diffusion. We may assume that the receiver can perfectly \emph{count} the number of molecules that fall into its sphere. This is the model used in \cite{PieAky11} (see also \cite{AkanNew}). The (microscopic) transparent receiver differs from the (macroscopic) sampling receiver as follows: as discussed in Section \ref{micro-section}, if a sampling receiver reads $\rho(x,t)$, a  transparent receiver reads a normal random variable whose mean is ${\rho}(x,t)$, and whose variance is ${\rho}(x,t)/V_R$.
\item \emph{Absorbing receiver:} The receiver absorbs any molecule that hits its surface. It keeps a count of the number of molecules that have hit it so far. Absorbing receivers imply the zero boundary condition when solving the Fick's law of diffusion.
\item \emph{Ligand or reactive receiver:} This model is based on the receptors of natural cells that are used in biological signaling pathways. It considers chemical kinetics of receptors located on the surface of the receiver. More specifically, it assumes that molecules reaching the surface of the receiver may react and bind with the receptors on the surface of the receiver and thereby initiate a chemical process inside the cell. The literatures generally consider cyclic adenosine monophosphate (cAMP) receptors that have a simple state space for each receptor: each receptor may be in the bound (B) or unbound (U) state with incoming molecules. Output at the receiver is the number of bound receptors. One important feature of a ligand receptor is its lingering effect, which is due to the fact that it takes some random time for each receptor to be detached, after binding. The state transition of each receptor depends on the concentration of molecules around the receptor, and is governed by chemical equations \cite{AkanReciever}.\\
The most simplifying model is to take the statistical average of the number of bound receptors to approximate the output, but then add a signal-dependent memoryless Gaussian noise to represent all the modeling imperfections, \emph{i.e.,} the output can be expressed as:\begin{align}Z=\alpha X+N\label{eqn:AtaAkaModel}\end{align} for some constant $\alpha$ and Gaussian noise $N$ \cite{AtaAka09}. In \cite{EinSarFekITW11}, a memoryless binomial$(k,p)$ distribution is used for the number of bound receptors, where $k$ is the total number of receptors and the binding probability $p$ is a function of concentration around the receptors. In \cite{Mahdavifar}, the number of bound receptors is modeled by a Poisson distribution whose parameter depends on the concentration of molecules around the receiver.\\
More accurate models of the ligand receptor use Markov chains to better represent chemical equations at the receptors and the memory of the system. The state transition probabilities of the Markov chain depend on the concentration of molecules around the receiver \cite{EinSarFekITW11, Ligand1}. More specifically, if there are $k$ receptors on the surface of the receiver, we can denote the state by a vector in $\mathbb{S}=\{B, U\}^k$. The state  at time instance $i$ is  also the output of the receiver, $Y_{i}=S_i$. The state transition probability $p(s_{i+1}|s_i, x_i)$ specifies the behavior  of the ligand receptor \cite{Ligand2}. The only assumption made about $p(s_{i+1}|s_i, x_i)$ in \cite{Ligand1} is that when a receptor is in bound (B) state, the probability that it becomes unbound (U) does not depend on concentration $X_i$.
\\
Finally, authors in \cite{ReactiveReceiver} consider the boundary condition that a Ligand receptor imposes (the boundary condition needed for solving the Fick's law of diffusion). The ligand receptor is modeled by a partially absorbing boundary condition, in conjunction with an active source of molecules. The partially absorbing boundary condition reflects the fact that molecules hitting a receptor might not bind with the receptor and get reflected back into the environment (hence partially absorbing, partially reflecting), and furthermore, a bound receptor might unbind and release a molecule and hence an active source of molecule is included in the model.
\end{itemize}

\subsection{End-to-end models}\label{sec:end-to-end-rev}
In this section, we review a number of existing end-to-end models derived by jointly considering the transmitter, diffusion channel  and receiver.  It may be impossible to analyze and model the end-to-end molecular channel independent from the release and reception mechanisms at the transmitter and receiver, respectively. In other words, the release and the reception mechanisms may have mutual effects with the molecular channel. Thereby, a joint analysis including the release, the transport, and the reception mechanisms is required.

Considering the three types of transmitter (timing transmitter, exact concentration and Poisson concentration transmitter) and the four types of receiver (sampling receiver, transparent receiver, absorbing receiver, and ligand receiver), one can potentially identify $3\times 4=12$ combinations. However, some of the combinations are not meaningful (\emph{e.g.}, the combination of a  timing transmitter and a sampling receiver), and some are mathematically challenging (\emph{e.g.} the combination of the ligand receiver with any of the three transmitters). While transparent receiver \cite{PieAky11,MahMakMou14,NoeCheSch14},  absorbing receiver \cite{YilHerTugCha14}, and ligand receptors \cite{PieAky11s,Cho15} are all considered in the literature, only  few of the possible transmitter/receiver combinations are studied. We shall review these combinations in the sequel.\footnote{See \cite{Heren, Akyldiz1} for two examples of models not discussed here.} In describing an end-to-end model, it is useful to keep in mind the
five types of noise that may need to be accounted for
\cite{NarimanSurvey}. The first type is the \emph{diffusion noise} due to the random propagation of individual molecules according to Brownian motion. We also have an environmental noise due to the \emph{degradation and/or reaction} of molecules. There is also a \emph{multiple transmitters noise} from molecules diffused by unintended transmitters. Finally, there are two types of noise due to the physics of transceivers: the \emph{transmitter emission noise} and \emph{receiver counting/reception noise}.

In all of the models presented here, for simplicity and for being tractable,  the transmitter is considered as a point source of molecules, located at the origin, and molecules are assumed not to change while propagating in the medium and also not interact with the transmitter. Motions of molecules do not influence each other and can be modeled by  independent processes. Also, when using a concentration transmitter, a time-slotted communications with symbol period of $T_s$ is assumed. The transmitter and receiver are assumed to be synchronized. Fick's laws are employed in the models in different ways to determine the parameters of the model.

\subsubsection{Exact concentration transmitter with sampling or transparent receivers}

The models that have an exact concentration transmitter with sampling or transparent receivers are also called the \emph{linear models} (also known as the \emph{deterministic models}).  They are described as follows:

\emph{Transmitter:} The transmitter is assumed to be completely controlling the intensity of molecules at its location. The transmitter releases a large number of molecules (macroscopic diffusion regime). The intensity should be non-negative.

\emph{Medium:} Extending the Fick's second law of diffusion (stated in \eqref{eqnFickrev21}) to three-dimensional space \cite{Akyldiz1} for a medium without any reaction and drift velocity, the  concentration of molecules in the environment, denoted by $\rho(\vec{r},t)$, at location $\vec{r}$ and at time $t$ is governed by
\begin{align}
\frac{\partial }{\partial t}\rho(\vec{r},t)=D\nabla^2\rho(\vec{r},t)+ c(\vec{r},t)\label{eqn:fick},
\end{align}
in which $D$ is the diffusion coefficient of the transmitted molecule, and $c(\vec{r},t)$ is the density of molecule production rate at point $\vec{r}$ at time $t$. Assuming that the transmitter is located at the origin, and at time $t=0$ suddenly releases one unit of molecules in the environment, the density of the molecule production rate will be equal to $c(\vec{r},t)= \delta(\vec{r}=0)\delta(t=0)$. In response to this input, the output $\rho(\vec{r},t)$ from equation \eqref{eqn:fick} is equal to the Green's function:
\begin{align}
h(\vec{r}, t)=\frac{\mathbf{1}[t>0]}{(4\pi Dt)^{1.5}}  e^{-\frac{\|r\|_2^2}{4Dt}}.\label{eqn:impulse}
\end{align}
Thus, \eqref{eqn:fick} describes a linear time-invariant system, whose impulse response is given in \eqref{eqn:impulse}.

The transmitter has a clock with frequency $f_s=1/T_s$, and instantaneously releases $X_{k}$ molecules every $T_s$ seconds, \emph{i.e.,} the density of production rate is the impulse train
$$c(\vec{r},t)= \sum_{k}X_k\delta(\vec{r}=0)\delta(t-kT_s),$$
then using the linearity of the diffusion system, the concentration of molecules at location $\vec{r}$ at time $t$ will be equal to
\begin{align}
\rho(\vec{r},t)=\sum\limits_{k}^{}X_{k}h(\vec{r}, t-kT_s)\label{eqnpXk}.
\end{align}

\emph{Receiver:}   The receiver has a clock too, with the same frequency $f_s$ (or possibly multiples of it), using it to uniformly sample the medium at times $jT_s$ for $j=0,1,2,\ldots$. We then have
\begin{itemize}
\item (Sampling receiver): assume that the receiver is modeled by a point, located at $\vec{r}^*$, and is not affecting the diffusion medium. Furthermore, we assume that the receiver can perfectly learn the macroscopic concentration of molecules at the time of sampling. Then, we obtain $Y_j=\rho(\vec{r}^*,jT_s)$. From \eqref{eqnpXk}, \begin{align}Y_j&=\sum\limits_{k}^{}X_{k}h(\vec{r}^*, jT_s-kT_s), \qquad j=0,1,2,\ldots
\\&=\sum\limits_{k}^{}X_{k}p_{j-k},\end{align}
has a convolution form where $p_j=h(\vec{r}^*, jT_s)$. From \eqref{eqn:impulse}, it is clear that $p_j=0$ for $j<0$, and thus the convolution can be written as:
\begin{align}Y_j&=\sum\limits_{k=0}^{\infty}p_k  X_{j-k}.\end{align}
\item (Transparent receiver): assume that the receiver is modeled by a transparent sphere of  volume $V_R$ rather than a point, but still not affecting the diffusion medium. Let us further assume that the receiver can perfectly \emph{count} the number of molecules that fall into its sphere. This is the model used in \cite{PieAky11} (see also \cite{AkanNew}). By similar arguments as given in Section \ref{sec:mic}, the distribution of the number of molecules falling in the receiver's volume at time $jT_s$ has a normal distribution whose mean is the macroscopic concentration $\sum\limits_{k=0}^{\infty}p_k  X_{j-k}$, and whose variance is $\frac{1}{V_R} \sum\limits_{k=0}^{\infty}p_k  X_{j-k}$. Thus, the total  number of molecules in receiver's sphere at time $jT_s$ satisfies
\begin{align}
Y_j  =\sum\limits_{k=0}^{\infty}p_k  X_{j-k}+N_j\label{eqn:impulse_dis},
\end{align}
where the noise, \emph{i.e.,}  $N_j$, is distributed according to $\mathcal{N}(0,\frac{1}{V_R} \sum\limits_{k=0}^{\infty}p_k  X_{j-k})$. Observe that when the input sequence (and thus $X_i$) is random, $N_j$ will be a doubly stochastic random variable since the variance of the $N_j$ is itself a random variable and depends on past inputs. This model is shown in Fig.~\ref{Linear}.

\begin{figure}
\begin{center}
\includegraphics[width=1\textwidth]{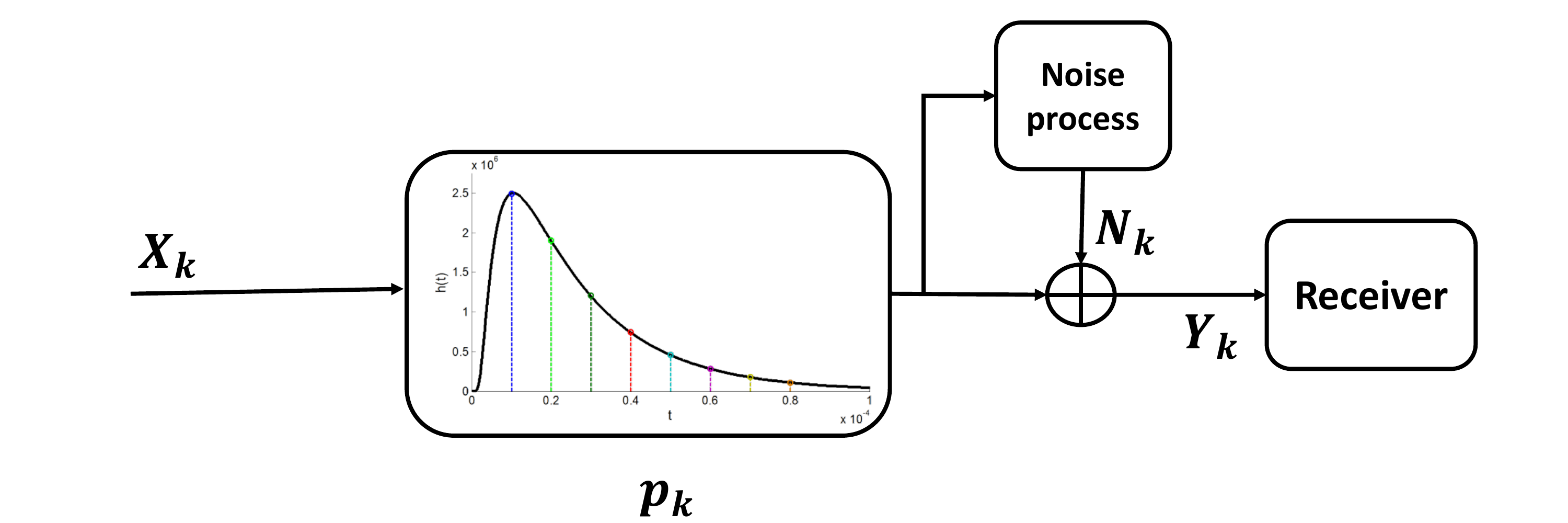}
\end{center}
\caption{The linear model (combination of the exact concentration transmitter and transparent receiver). It takes the input sequence $X_i$ and convolves it with samples of the impulse response of the diffusion channel. Finally, a signal dependent noise is added to the convolution term. }\label{Linear}
\end{figure}

\begin{remark}\label{remark2d}
Equation \eqref{eqn:impulse_dis} only provides the pmf (probability mass function) of $Y_j$ given the input sequence. To completely specify the channel, one has to specify the correlation of $Y_j$ and $Y_{j'}$ (for any given $j,j'$), conditioned on the input sequence. In other words, the correlation of $N_j$ and $N_{j'}$ conditioned on $X_0, X_1, \ldots$ should be specified. Observe that the number of molecules in $V_R$ is changing continuously over time, increasing or decreasing by one when a molecule enters/exits the receiver volume. Therefore, for very small values of $T_s$, $Y_j$ and $Y_{j+1}$ will be correlated, conditioned on the past inputs, and we cannot impose independence on $N_j$ and $N_{j+1}$.\footnote{  However, one can approximate $N_j$'s with jointly Gaussian colored noises, whose correlation coefficients and variances depend on the input sequence $X_0, X_1, \ldots$. Briefly speaking, the correlation of $N_j$ and $N_{j+1}$ can be computed as follows: let $q_{11}$ be the probability that a released molecule falls in the receiver's sphere at times $jT_s$ and $(j+1)T_s$. Let $q_{10}$ be the probability that the molecule falls in the sphere at time $jT_s$, but does not fall at time $(j+1)T_s$; $q_{01}$ is defined reversely and $q_{00}$ is the probability that molecule does not fall in either of times $jT_s$ or $(j+1)T_s$. If we release a deterministic number $x$ of  molecules at time zero, we can define a multinomial distribution $Z_{00}, Z_{01}, Z_{10}, Z_{11}$ with parameters $x$ and probabilities $(q_{00}, q_{01}, q_{10}, q_{11})$. Then $Y_j=Z_{10}+Z_{11}$ and $Y_{j+1}=Z_{01}+Z_{11}$ are the number of molecules received in times slots $jT_s$ and $(j+1)T_s$ respectively. If $q_{01}, q_{10}, q_{11}$ are small, we can approximate $Z_{10}, Z_{01}$ and $Z_{11}$ by independent Gaussian distributions. Thus, $Y_j$ and $Y_{j+1}$ will be jointly (colored) Gaussian random variables when we transmit $x$ molecules at time zero.} This analysis is missing in the literature \cite{PieAky11, AkanNew}.
\end{remark}
\end{itemize}

From the five types of noise mentioned in Section~\ref{sec:end-to-end-rev}, none of them was considered in the model given for the sampler receiver. And only diffusion noise was considered in the model given for the transparent receiver. It is possible to modify the model to consider other noises in the model. For instance consider the case in which the receiver has a small volume $V_R$ and \underline{imperfectly} counts the number of molecules that fall into that area. This incurs a \emph{particle counting noise} in addition to the diffusion noise.

\subsubsection{Poisson concentration transmitter with absorbing receiver}

Poisson concentration transmitter with absorbing receiver is also called the \emph{Poisson Model}.

\emph{Transmitter:}
The transmitter chooses a rate $X_i\geq 0$ at time slot $i$.
The number of released molecules in the beginning of the $i$-th time-slot  is a Poisson random variable, where its average or rate, is $X_i$. As a result, the density of production rate  is the impulse train
$$c(\vec{r},t)= \sum_{k}S_k\delta(\vec{r}=0)\delta(t-kT_s),$$
where $S_k\sim\mathsf{Poisson}(X_i)$ is the number of released molecules at time $kT_s$ by a transmitter located at the origin. Random variable $S_k$ is a  doubly stochastic random variable (a Poisson distribution whose  parameter is the random variable $X_i$). 

\emph{Medium and receiver: }
Each released molecule is absorbed upon hitting the receiver.
Let $p_k$, $k=0,1,2,\ldots$, denote the probability that a released molecule at the current slot hits the receiver in the next $k$-th time slot. The values of $p_k$'s depend on the communication medium and in general can be derived from Fick's diffusion law. For a one-dimensional motion, the distribution of the hitting time $T$  was found in Example \ref{example:rev1}. Then, the hitting probability can be computed as follows: \begin{align}p_k=\int_{kT_s}^{(k+1)T_s}f_T(t).\label{pkdefrev}\end{align}
Distribution of the first hitting time (and the values of $p_k$) for a  non-uniform medium that is arbitrarily filled with barrier or obstacles may be found by numerically solving the Fick's equations.

Generally, the signal is decoded at the receiver based on the total number of molecules received during the individual time slots. In a time slot, three sources contribute to the received molecules: (i) molecules due to the transmission in the current time slot, (ii) the residue molecules due to the transmission in the earlier time slots known as the interference signal, and (iii) noise molecules, modeled within a slot of duration $T_s$ as a Poisson random variable with parameter $\lambda_0 T_s$.
Based on the thinning property of Poisson distribution, the number of received molecules in the time slot $i$, denoted by $Y_i$, is as follows \cite{Ourwork1}:
\begin{align}
Y_i\sim\Po\left(\lambda_0T_s+\sum\limits_{k=0}^{i}p_{k} X_{i-k}\right)=\Po(\mathtt I_i+p_0X_i)\label{eqn1nn}
\end{align}
where $\mathtt I_i$  is the sum of interference and noise at time $i$. The above equation states $p(y_i|x_{[0:i]})$, which together with \begin{align}p(y_{[0:n]}|x_{[0:n]})=\prod_{i=1}^n p(y_i|x_{[0:i]})\label{eqn:newadd1}\end{align} describes the Poisson channel completely. Equation \eqref{eqn:newadd1} is proved using the thinning property of Poisson distribution in \cite{Ourwork1}.

From the five types of noise mentioned in Section~\ref{sec:end-to-end-rev}, transmitter noise, diffusion noise and multiple transmitters noise are taken into account. Transmitter noise is reflected in the fact that the number of molecules exiting the transmitter cannot be exactly controlled; diffusion noise is considered because the model is based on microscopic Brownian motion, and finally multiple transmitters and background noise is considered by the $\lambda_0T_s$ term.

\subsubsection{Timing transmitter with absorbing receiver}
Timing transmitter with absorbing receiver leads to what is known as the \emph{Timing Model} for MC.

\emph{Transmitter:} The transmitter is assumed to be completely controlling the release time of molecules one by one at its location. Information is coded in the release time of individual molecules \cite{Timing1}.

\emph{Medium and receiver:} The released molecules randomly propagate according to the Brownian motion. Each molecule may arrive at the receiver after a (random) delay, or may never arrive at the receiver.\footnote{The lifetime of a molecule may be modeled by Weibull distribution \cite{Arifler} which is a widely used lifetime distribution, or bounded by a hard threshold \cite{Timing4n}.} Molecules do not arrive if the Brownian motion is transient, or the molecules fade away in the environment. Since we consider an absorbing receiver,   each molecule hits the receiver at most once. Despite the uncertainty in the arrival times of the molecules, these arrival times are statistically correlated with their release time, and this correlation can be used for information transmission. In this context, information transmission capacity may be measured by bits per unit of time \cite{Timing4n, Timing2/52}, bits per molecule \cite{Timing2/52}, or bits per joule \cite{Timing6} for a given production rate of molecules.

Timing channel models generally ignore the existence of noise molecules, \emph{i.e.,} molecules of the same type in the environment that are not diffused by the transmitter. Noise molecules are likely to complicate the problem significantly. From the five types of noise mentioned in Section~\ref{sec:end-to-end-rev}, only diffusion noise is taken into account.\footnote{Some variations of the timing model consider a maximum lifetime for molecules; hence a form of degradation noise is considered as well.}

Assuming that a molecule is released at time $X$, the arrival time is equal to $Z=X+T$, where $T$ is the transmission delay. Because molecules are absorbed upon hitting the receiver, the distribution of $T$ is that of a first arrival time.  For a one-dimensional motion with variance $\sigma^2$, a positive drift $v>0$ and travel distance $d$, $T$ will follow an inverse Gaussian distribution $\ig(\mu,\lambda)$ given in equation \eqref{eqn:IG}. When there is no drift $v=0$, $T$ follows the L\'evy distribution given in equation \eqref{eqn:Levy}.

\subsubsection{Ligand receiver with simplification on the transmitter side}\label{sec:ligand-model-part}

 These models focus on the physical complexity of the receiver and simplify the channel noise, with the goal of understanding fundamental limits imposed by the receiver alone.

\emph{Transmitter and medium:} It is assumed that the concentration of molecules around the receiver at time slot $i$ is a function of transmission in that time slot, \emph{i.e.,} $f(X_i)$ for some known function $f(\cdot)$.

\emph{Receiver:} A receiver with cAMP receptors that was described in Section \ref{sec:rece:rev} is assumed.

\section{Literature review: concentration transmitter}\label{sec:concen}

One of the popular signaling  methods in molecular communication uses the molecules' concentration to encode the information. In this setup, the transmitter encodes information into the concentration of released molecules. The released molecules follow a diffusion process through the channel to reach the receiver. As discussed in Section~\ref{sec:model}, the temporal and spatial concentration of molecules can be derived by the Fick's second law of diffusion which results in a distance-time dependent impulse response.

There are several works in the literature on the concentration signaling, many of which claim complete characterizations of channel capacity, leaving the general impression to unfamiliar readers that the capacity of molecular channels is solved. However, once we go inside the papers we see that the capacity calculations are mostly done after  simplifications in terms of the  channel memory. Simplification and approximation have the advantage of leading to explicit  expressions. But one needs to also address how the derived bounds relate to the actual channel capacity without any simplifications, by specifying whether they are lower or upper bounds to the actual capacity.

Having said that, in principle, the capacity of concentration signaling, considering the transmitter-receiver limitations and the intersymbol interference (ISI) effect of the channel is not an open problem, under some reasonable physical assumptions. As pointed out in \cite{AmiArjGohNasMit15}, we may model the diffusion channel (possibly with drift in a non-uniform medium) as a state dependent channel. Here the state models the number or density of molecules across the environment; we can divide the space into very small cells and keep the number of molecules in the cells as the current state of the medium. Part of the state can also model the bound/unbound state receptors on the surface of the receiver. Then, observe that the resulting state-dependent channel is \emph{indecomposable} \cite[p. 105]{Gallager} as the initial state \emph{diffuses away over time} under reasonable physical constraints on the medium. Thus, one can characterize the capacity in a computable form \cite{Gallager}. However, the state space is very large and capacity characterization is in finite-letter form, making it prohibitively hard to compute and thus of little practical value.\footnote{From an information theoretic perspective, a characterization of the capacity region is called computable if there exists an algorithm (constructed based on the characterization) that gets an approximation level $\epsilon$ as its inputs and halts in \underline{finite time} and produces a curve within $\epsilon$ distance of the capacity region. There is no restriction on the time that takes for the algorithm to halt.}

 In the following, we describe some approaches that exist in the literature.

\subsection{Linear Model}

The goal is to study the capacity of the channel
\begin{align}
Y_j  =\sum\limits_{k=0}^{\infty}p_k  X_{j-k}+N_j, \label{eqn:pkXNkj}
\end{align}
where the noise, \emph{i.e.,}  $N_j$, is distributed according to $\mathcal{N}(0,\frac{1}{V_R} \sum\limits_{k=0}^{\infty}p_k  X_{j-k})$. The input constraint consists of non-negativity of input concentration $X_i\geq 0$, and possibly maximum and average concentration constraints: $|X_i|\leq \mathsf{A}$ and $\sum_{i=1}^nX_i/n\leq \mathcal{E}_{\textnormal{s}}$. To study the capacity of this channel, one has to specify the joint distribution of $N_1, N_2, \ldots$ conditioned on the entire input sequence as discussed in Remark \ref{remark2d}. Even with the assumption of conditional independence of $N_j$ (which is justified when $T_s$ is large), as far as we are aware, the capacity of the above channel has not been studied. If the variance of $N_j$ were some constant $\sigma^2$ and channel inputs were allowed to become negative, the problem would have reduced to the classical Gaussian ISI channel that has been subject to many studies, starting from the work by Hirt and Massey in \cite{Massey}.  Nonetheless, much of the classical ideas clearly carry over from the classical ISI channels. For instance, one can find bounds on capacity using the ideas of using i.i.d.\ input distribution \cite{Shamai}. Also, when the number of terms in the linear expansion \eqref{eqn:pkXNkj} is finite ($p_k>0$ for large enough $k$), we can find computable finite-letter expressions for the capacity using the ideas presented later in Section \ref{sec:PoissonModel}.

The existing literature on the linear model simplify the problem by discarding the channel memory. In this approach, the molecular channel is approximated by a memoryless channel (in many cases, a channel with binary input alphabet) whose capacity can be readily found by maximizing the input-output mutual information (\emph{e.g.,} see \cite{AtakanAKan08, LiuYan14, AtaAka09}). Thus, while these papers might be valuable in the context of modeling, they are not exciting to information theorists. These approaches do not genuinely consider a channel with memory, as the interference from past inputs (ISI) is either ignored, or averaged and put into the transition probabilities of a memoryless channel.\footnote{A variation of  this approach is proposed in \cite{Kuran10} by considering two binary channel, one of which occurs depending on the last transmitted symbol.
}

The binary input restriction may be justified by the fact that nano nodes should be simple, and one of the simplest transmission strategies is the On-Off Keying modulation in which no molecule is transmitted for information bit 0 and $A_{\max}$ concentration is released for information bit 1. While On-Off keying simplifies the transmitter's design, receiver's design can be simplified to a simple threshold decoder by adopting transmission modulation schemes that reduce or mitigate the ISI. Thus, channel simplifications and restriction to certain modulation schemes may not be unjustified as it may appear in first glance.

Finally, there are also some works based on the \emph{quorum sensing} property of bacterial colonies \cite{EinSarFekISIT12,EinSarFekTWC13} that also consider a memoryless channel, but with an input-dependent noise. Even though we obtained the linear model for an exact concentration transmitter, the same end-to-end model applies to the works based on quorum sensing.\footnote{By ignoring ISI and considering $N_j$ as the transmitter or the receiver noise in \eqref{eqn:pkXNkj}, the linear model reduces to the model of these works (which was obtained by using a Gaussian approximation for the binomial distribution) if the dependency between the transmitter and receiver noises is ignored.}
In these works, the transmitter and receiver employ bacterial colonies. The collective behavior  of the bacteria in response to stimuli is exploited for transmission and reception. The colony on the transmitter side, in the steady state, senses the concentration of a special molecule type and in response produces and releases another type of molecules. The released molecules propagate in the medium based on diffusion equations, which has been considered in the steady state. Similar to the transmitter, the receiver senses the concentration based on the quorum sensing property. All limitations, \emph{i.e.,}  noises, are modeled with an additive Gaussian noise with input-dependent variance (still a memoryless channel), and no ISI terms are considered. In the same scenario of biological nodes consisting of bacteria, the relaying is studied in \cite{EinSarFekISIT13, EinSarFekICC14}, where the relay exploits its quorum sensing property to apply the sense and forward scheme (which is parallel to the amplify and forward scheme in the classic communication).

\subsection{Poisson Model}\label{sec:PoissonModel}
The goal is to compute the capacity of the LTI-Poisson channel
\begin{align}\label{eqn:newrevpoisson1}
Y_i\sim\Po\left(\lambda_0T_s+\sum\limits_{j=0}^{i}p_{j} X_{i-j}\right)
\end{align}
under average and maximum intensity cost constraints. More specifically,   the following constraints on the input codewords of length $n$ are assumed: $X_i\geq 0$, average input constraint $\sum_{i=1}^nX_i/n\leq \mathcal{E}_{\textnormal{s}}$ and a constraint on the maximum value of $X_i$, $X_i\leq \mathsf{A}$.

The summation $\sum_{j=0}^ip_{j}X_{i-j}$ is the convolution of the sequence $\left(X_0, X_1,\ldots\right)$ with the sequence $\textbf{p}=\left(p_0, p_1, p_2, \ldots\right)$. This makes the $Y_i$ as the output of a channel consisting of a cascade of an LTI system (with impulse response $\textbf{p}$) and a memoryless Poisson channel, called an \emph{LTI-Poisson channel} in \cite{AmiArjGohNasMit15}. This model can be understood as a generalization of the classical memoryless Poisson channel. Therefore, the LTI-Poisson model relates to two bodies of literature in information theory: networks with memory and memoryless Poisson channels. A common point in both literatures is an attempt to find easy-to-compute expressions for the capacity (e.g. see \cite{Lapidoth2011, LapidothMoser2009}).

The capacity for the channel given in equation \eqref{eqn:newrevpoisson1} is claimed to have been solved in \cite[p. 9, eqn. (40)]{AkyldizPoisson}. However, in the proof on page 10 of \cite{AkyldizPoisson}, after equation (50), it is claimed that $Y_i$'s are i.i.d.\ and $H(Y^n)$ is expanded as $\sum_{i=1}^n H(Y_i)$. But $Y_i$'s are correlated in general because they depend on the input sequence $X_{i}$.

The capacity for this channel has been characterized in \cite{AmiArjGohNasMit15} under the assumption that molecules injected into the environment will disappear after $k$ time-slots, for some large enough $k$. Thus, $p_i=0$ for $i>k$. This allows one to write that
\begin{align}
Y_i\sim\Po\left(\lambda_0T_s+\sum\limits_{j=0}^{k}p_{j} X_{i-j}\right).
\end{align}
In particular, $p(y_i|x_{[0:i]})=p(y_i|x_{[i-k:i]})$, and from equation \eqref{eqn:newadd1}
\begin{align}p(y_{0:n}|x_{1:n})=\prod_{i=0}^np(y_i|{x}_{i:i-k}).\label{MLC}
\end{align}
The factorization given in equation \eqref{MLC} allows for a multi-letter, albeit computable characterization of the capacity region. Factorization of the type given by \eqref{MLC} was first exploited by Verdu for MAC channels \cite{Verdu}. Verdu's motivation for defining this class of networks was to study  linear ISI channels. The intuitive reason that factorization of \eqref{MLC} is useful is that one can set or reset the channel state using any $k$ consecutive inputs (as the channel remembers only the last $k$ inputs)\cite{AmiArjGohNasMit15}.

In \cite{AmiArjGohNasMit15}, the capacity of the original channel (with memory) is sandwiched between the capacities of two memoryless channels, \emph{i.e.,} two memoryless channels are given whose capacities bound the desired capacity from below and above. The upper and lower bounds can be made arbitrarily close to each other, resulting in a computable characterization of the capacity region.

For the lower bound, a natural number $r$ is chosen. Then time is partitioned into  blocks of size $k+r$, \emph{i.e.} one block for time instances 1 to $k+r$, one block for time instances $k+r+1$ to $2(k+r)$, etc. Then, the channel is
depreciated by deleting output ${Y}_i$'s for the first $k$ time instances of each block. In other words, the new channel after deletion has  inputs ${X}_1, {X}_2, \cdots$ and outputs ${Y}_{k+1}, {Y}_{k+2}, \cdots, {Y}_{k+r}$ and then ${Y}_{2k+r+1}, {Y}_{2k+r+1}, \cdots, {Y}_{2k+2r}$, etc. Because the outputs in each block depend only on inputs in the same block, the resulting channel is \emph{block memoryless} and its capacity lies below the capacity of original channel.

For the upper bound,  a natural number $r$ is chosen.  Then time is partitioned into  blocks   of size $r$; in other words, first block covers time instances 1 to $r$, second block covers time instances $r+1$ to $2r$, etc. The channel is enhanced by allowing the transmitter to arbitrarily ``reset" the memory of the channel ($k$ last inputs) at the \emph{beginning} of each block (without any regard to its actual last $k$ inputs). The new channel has a higher capacity than the original channel, since the memory content specified by the transmitter at the beginning of each block can simply be the actual state that the system would have been in, if transmitter did not have the option of changing the memory content of the channel. Furthermore, the new channel is \emph{block memoryless} and its capacity lies above the capacity of original channel.

\textbf{The symmetrized Kullback-Leibler divergence upper bound:} Capacity of a memoryless channel can be characterized as the maximum over input distributions of the mutual information between channel input and output of the channel. This characterization is not always sufficiently explicit.  Another contribution of \cite{AmiArjGohNasMit15} is to propose an easy to compute and explicit upper bound on mutual information.

The symmetrized Kullback-Leibler divergence (KL divergence) is defined as $D_{\mathsf{sym}}(p\|q)=D(p\|q)+D(q\|p)$. Then,
\begin{align}
D_{\mathsf{sym}}(p(x,y)\|p(x)p(y))&=D(p(x,y)\|p(x)p(y))+D(p(x)p(y)\|p(x,y))\nonumber
\\&\geq D(p(x,y)\|p(x)p(y))\nonumber\\&=I(X;Y).\label{jensen}
\end{align}
One can prove \eqref{jensen} directly by first simplifying the expression of $D_{\mathsf{sym}}(p\|q)$ and then applying the Jensen's inequality \cite{AmiArjGohNasMit15}, but the above chain of inequalities illustrate that the gap $D_{\mathsf{sym}}(p(x,y)\|p(x)p(y))-D(p(x,y)\|p(x)p(y))$ is $D(p(x)p(y)\|p(x,y))$, the \emph{lautum information} (``mutual" written in the reverse order) that is an object of interest on its own terms \cite{VerduLautum}.

Given a channel $p(y|x)$, one can then define the following upper bound on capacity \cite{AmiArjGohNasMit15}:
\begin{align}\label{KL}
\mathcal{U}(p(y|x))&=\max_{p(x)}D_{\mathsf{sym}}(p(x,y)\|p(x)p(y))\geq \max_{p(x)}I(X;Y)=C(p(y|x)).
\end{align}
It is shown in  \cite{AmiArjGohNasMit15} that $\mathcal{U}(p(y|x))$ can be explicitly computed. For instance, for a point to point Poisson channel $p(y|x)$, where $Y=\mathsf{Poisson}(X+\lambda_0)$, it has the following compact formula:
\begin{equation}
I(X;Y)\leq \mathcal{U}(p(x,y))=\mathsf{Cov}(X+\lambda_0, \log(X+\lambda_0)),
\end{equation}
where $\mathsf{Cov}(X,Y)=\mathbb{E}[XY]-\mathbb{E}[X]\mathbb{E}[Y]$.
Furthermore, it yields previously unknown bounds for channels with small capacity (which can occur in MC systems). For instance, for a Poisson channel with average intensity constraint $\mathcal{E}_{\textnormal{s}}$ and maximum intensity constraint $\mathsf{A}$, this bound is calculated as:
\begin{align*}
&\mathcal{U}_{\mathsf{Poisson}}\big(p(y|x)\big):=\max_{\substack{p(x):\\ \mathbb{E}[X]=\mathcal{E}_{\textnormal{s}},~~ 0 \leq X \leq \mathsf{A}}}\mathcal{U}(p(x,y))=\\
&\quad \begin{cases}\frac{\mathcal{E}_{\textnormal{s}}}{\mathsf{A}}(\mathsf{A}-\mathcal{E}_{\textnormal{s}})\log\left(\frac{\mathsf{A}}{\lambda_0}+1\right), & \mathcal{E}_{\textnormal{s}}< \mathsf{A}/2
\\
\frac{\mathsf{A}}{4}\log\left(\frac{\mathsf{A}}{\lambda_0}+1\right), & \mathcal{E}_{\textnormal{s}}\geq \mathsf{A}/2.
\end{cases}
\end{align*}
For previous works on Poisson channel and other techniques for bounding mutual information, see \cite{Lapidoth2011, LapidothMoser2009}.

\subsection{Other models}
Authors in \cite{Ghavami} consider a combination of an exact concentration transmitter and an absorber receiver. The exact concentration transmitter can send $0\leq X_i\leq N$ molecules at the beginning of each time slot. 
The receiver counts the number of absorbed molecules in each time slot. Consider the hitting probabilities $\{p_k\}$ of the absorbing receiver in equation \eqref{pkdefrev}. Then, each of the $X_0$ molecules released in the first time slot arrive in the $k$-th time slot with probability $p_k$. Thus, the total number of molecules that are released in time slot $0$ and arrive in  the $k$-th time slot follows a binomial distribution with parameters $(X_0, p_k)$. Since  we have a transmission at the beginning of each time slot, the total number of molecules received at the time slot $k$ is the sum of independent binomial random variables corresponding to transmissions from time slots $k$, $k-1$, $k-2$, etc. The sum of independent binomial variables does not have a nice analytical form. The more serious difficulty is the correlation that arises between outputs at different time-slots.\footnote{This is the main advantage of the Poisson concentration transmitter over the exact concentration transmitter in the presence of an absorbing receiver (see equation \eqref{MLC}). The linear model (that uses a transparent receiver) also has a similar disadvantage as discussed in Remark \ref{remark2d}.} There are only some approximation techniques for handling dependencies that arise in such balls and bins problems; see for instance \cite[Section 5.4]{Mitzenmacher}. Authors in \cite{Ghavami} simplify the problem by assuming that $p_k=0$ for $k\geq 2$. Furthermore, they handle the correlation by assuming an i.i.d.\ input distribution when evaluating the $n$-letter mutual information form of the channel capacity. We will review the idea of evaluating the $n$-letter mutual information by assuming an i.i.d.\ distribution in relation to a different problem in detail in Section \ref{sec:rel_time}.

A different model for the diffusion medium is considered in \cite{EinSarBeiFekISIT11}. The transmitter is still an exact concentration transmitter, but instead of using the linear model and Fick's law to evaluate the concentration at the receiver, the authors assume that the concentration at the receiver can be at either of the two ``low" or ``high" concentration states. In other words, the communication medium is modeled with a two-state Markov chain, with low and high states, which indicate the ``overall" intensity of residual molecules in the environment due to the past transmissions (high ISI or low ISI).  By using the On-Off Keying scheme and assuming memory of depth one, an achievable rate is derived in \cite{EinSarBeiFekISIT11}.

\section{Literature review: timing transmitter}\label{sec:rel_time}

The capacity of molecular timing channels has been the subject of several studies. These works appeal to information theorists because they exploit serious information theoretic tools and are mathematically rigorous. While it is not possible to reproduce the entire literature here, we selectively provide a rough sketch of some of the tools and the ideas used.

\textbf{Signaling  with identical tokens:}
In an early work \cite{Timing1}, the author assumes that $n$ molecules are released at times instances $x_1, x_2, \cdots, x_n$, and arriving at the receiver at times $z_1=t_1+x_1, z_2=t_2+x_2, \cdots, z_n=t_n+x_n$, where $t_i$ is the travel time of the $i$-th molecule. Travel time $T_i$ are assumed to be independent and identically distributed according to some smooth continuous  density function (the inverse Gaussian or L\'evy distribution for free diffusion in a one-dimensional medium).  While the transmitter and receiver are perfectly synchronized, the order according to which the particles are received is not necessarily the same as the order they are released. The molecules are of the same type and indistinguishable at the receiver, which is the main source of difficulty in the problem. Therefore, the receiver has only the sorted values $\mathsf{sort}(z_1, z_2, \cdots, z_n)$, not the exact vector $(z_1, z_2, \cdots, z_n)$. This is a main difference of this model with ``bits through queues" of \cite{Venkat}. Here we are interested in the mutual information $$I(X^n; \mathsf{sort}(Z^n)).$$
This mutual information is harder to analyze than
$I(X^n; Z^n)$ as in \cite{Venkat}. The reason is that \cite{Timing2/5}
$$I(X^n; Z^n)=h(Z^n)-h(Z^n|X^n)=h(Z^n)-h(T^n)$$
where $h(\cdot)$ is the differential entropy. Here  only the first term $h(Z^n)$ depends on input pmf $p(x^n)$. On the other hand,
$$I(X^n; \mathsf{sort}(Z^n))=h(\mathsf{sort}(Z^n))-h(\mathsf{sort}(Z^n)|X^n)$$
but $h(\mathsf{sort}(Z^n)|X^n)$ depends on $p(x^n)$. However, we have that  \cite{Timing2/5}  \footnote{If
$Z_\Delta^n$ is the quantized version of $Z^n$ (a discrete random variable), we get that $H(Z_\Delta^n|X^n)=H(\mathsf{sort}(Z_\Delta^n), Z_\Delta^n|X^n)= H(\mathsf{sort}(Z_\Delta^n)|X^n)+H(Z_\Delta^n|\mathsf{sort}(Z_\Delta^n), X^n)$. Letting $\Delta$ converge to zero, the difference $H(Z_\Delta^n|X^n)- H(\mathsf{sort}(Z_\Delta^n)|X^n)$ converges to
$h(Z_\Delta^n|X^n)- h(\mathsf{sort}(Z_\Delta^n)|X^n)$, and $H(Z_\Delta^n|\mathsf{sort}(Z_\Delta^n), X^n)$ converges to $H(Z^n|\mathsf{sort}(Z^n), X^n)$.
}
$$h(Z^n|X^n)=h(\mathsf{sort}(Z^n)|X^n)+H(Z^n|\mathsf{sort}(Z^n), X^n).$$
Thus, $h(\mathsf{sort}(Z^n)|X^n)=h(T^n)-H(Z^n|\mathsf{sort}(Z^n), X^n)$. Hence,
$$I(X^n; \mathsf{sort}(Z^n))=h(\mathsf{sort}(Z^n))+H(Z^n|\mathsf{sort}(Z^n), X^n)-h(T^n).$$
Since the third term on the right hand side $h(T^n)$ does not depend on $p(x^n)$, one needs to consider the maximum of sum of the first two terms as a function of $p(x^n)$. Authors in \cite{Timing2/5} proceed by finding upper bounds on these two terms and maximize those bounds over input pmfs $p(x^n)$. It is also worth to mention another idea of this paper: observe that $\mathsf{sort}(Z^n)$ does not depend on the order of $Z_1, \ldots, Z_n$. Therefore if $\pi$ is a permutation from $\{1,2,\cdots, n\}$ to itself, then
$$\mathsf{sort}(Z_1, Z_2, \ldots, Z_n)=\mathsf{sort}(Z_{\pi_1}, Z_{\pi_2}, \ldots, Z_{\pi_n}).$$ Thus, if instead of transmitting the input sequence in the order of $X_1, X_2, \ldots, X_n$, we transmit them in the order of $X_{\pi_1}, X_{\pi_2}, \ldots, X_{\pi_n}$, the mutual information between $X^n$ and $\mathsf{sort}(Z^n)$ would not change. In other words, if $p(x^n)$ is a capacity achieving distribution and we define $\tilde{X}_i=X_{\pi_i}$, then $p(\tilde x^n)$ would also be a capacity achieving distribution. Observing that mutual information is concave in the input distribution, $q(x^n)=\frac{1}{2}p_{X^n}(x^n)+\frac 12 p_{\tilde{X}^n}(x^n)$ would also be capacity achieving. If we consider all the permutations $\pi$, and take their average pmf, we get that it suffices to take maximum over input distributions that are symmetric with respect to permutation on the input indices (called ``hypersymmetric" in \cite{Timing2/5}).

We continue this part by reviewing some of the ideas of \cite{Timing1, Timing2}: Author in \cite{Timing1} uses the following idea for numerically computing lower bounds on the mutual information $I(X^n; \mathsf{sort}(Z^n))$.
Suppose we have an intractable channel $p(y|x)$ and an input distribution $p(x)$.
The idea is to find a tractable approximation $q(x|y)$ of the channel $p(x|y)$.  For any arbitrary channel $q(y|x)$ we have that
$$I(X;Y)=\mathbb{E}_{p(x,y)}\log \frac{p(x|y)}{p(x)}\geq \mathbb{E}_{p(x,y)}\log \frac{q(x|y)}{p(x)},$$
where $q(x|y)$ is calculated from $q(x,y)=p(x)q(y|x)$. The above inequality can be directly verified. Note that similar upper bounds on mutual information can be found via Topsoe's inequality \cite{Topsoe}: for any arbitrary output pmf $q(y)$ we have
\begin{align}I(X;Y)=\mathbb{E}_{p(x,y)}\log \frac{p(y|x)}{p(y)}\leq \mathbb{E}_{p(x,y)}\log \frac{p(y|x)}{q(y)}.\label{topsoe}\end{align}

\textbf{Memoryless models:} As we saw, molecules arriving out of order are a challenge. This might be avoided if we release molecules of different types so that they can be distinguished at the receiver \cite{Timing3}.  Alternatively if we assume a lifetime for molecules (after which they fade away in the environment), we might restrict ourselves to a certain class of encoders that after releasing a molecule, delays the next transmission long enough to ensure that the previous transmission has either hit the receiver or died out in the environment \cite{Timing4n}.\footnote{This idea resembles the interference mitigation modulations techniques for dealing in channels with intersymbol interference (ISI). } Even though this restriction may not optimal, but by studying the maximum achievable rate for this class of encoders, we can find lower bounds on the capacity of the molecular timing channel.

Having resolved the out of order problem, the capacity of the molecular timing channel reduces to the capacity of an additive channel $Z=X+T$. Distribution of the transmission delay $T$ depends on the physical properties of the communication medium. For a one-dimensional motion with variance $\sigma^2$ and a positive drift  velocity $v$ towards the receiver located at distance $d$ from the transmitter, $T$ will follow an inverse Gaussian distribution $\ig(\mu,\lambda)$ given in equation \eqref{eqn:IG}. The additive channel $Z=X+T$ is called an AIGN (additive inverse Gaussian noise) channel. If there is no drift towards the receiver $v=0$, the channel becomes an ALN (additive L\'evy noise) channel \cite{Timing4n}. 

 The  goal is to compute the following expression:
$$\max_{p(x): X\geq 0, \mathbb{E}[X]\leq \Lambda} I(X;X+T),$$
for some $\Lambda>0$.
Even though capacity is concave in the input density and convex optimization techniques may be employed,  the space of distributions over which the maximization occurs is difficult to handle analytically. There are several works that find explicit lower, upper or asymptotic expressions for the capacity of AIGN under an average input cost constraints \cite{Timing3, Timing4, Timing5}, with some of the techniques are borrowed (but carefully adapted) from earlier works on the Poisson channel. A simplifying fact is the
additivity property of the IG distribution \cite{Timing3}:
let $T_1\sim
\ig(\mu_1,\lambda_1)$ and $T_2\sim
\ig(\mu_2,\lambda_2)$ be  independent
 random variables. We further assume that $\frac{\lambda_1}{c_1\mu_1^2}=\frac{\lambda_2}{c_2\mu_2^2}=\kappa$ for some $c_1, c_2>0$. Then, $c_1 T_1+c_2 T_2\sim \ig(c_1\mu_1+c_2\mu_2,\kappa(c_1\mu_1+c_2\mu_2)^2)$.

In the following we mention some of the proof ideas. Observe that $I(X;X+T)=h(X+T)-h(T)$. Therefore, the problem is to maximize $h(X+T)$ subject to $X\geq 0, \mathbb{E}[X]\leq \Lambda$ for some $T\sim \ig(\mu,\lambda)$.
\begin{itemize}
\item (Lower bounds:) The lower bound in \cite{Timing3} is derived by evaluating $h(X+T)$ when $X$ is distributed according to an IG distribution (capacity is maximum over all distributions, so this yields an inner bound). The above additivity property of IG distribution is used to specify the distribution of $X+T$. To derive a lower bound in \cite{Timing4}, author chooses the input distribution of $X$ to an exponential distribution. The distribution of $X+T$ is the convolution of an exponential and an IG distribution. The author avoids this calculation. Instead  $h(X+T)$ is bounded from below using the entropy power inequality in terms of $h(X)$ and $h(T)$. In \cite{Timing5}, the exact formula of the convolution of an exponential and an IG distribution is cited from \cite{Timing5n} and a new analytical lower bound is derived.
\item (Upper bounds:) The upper bound in \cite{Timing3} is derived by noting that $\mathbb{E}[X+T]=\mathbb{E}[X]+\mathbb{E}[T]\leq \Lambda+\mu$. Thus, the entropy of $X+T$ is bounded by the entropy of the exponential distribution with mean $\Lambda+\mu$, as exponential distribution has maximal differential entropy amongst all non-negative random variables with the same mean. The idea of \cite{Timing5} is to use Topsoe's inequality in \eqref{topsoe} to bound mutual information from above, with the choice of inverse Gaussian distribution for output pmf.
\end{itemize}
In \cite{Timing4n}, authors  consider a diffusion medium with no drift. A lifetime for molecules is considered (after which they die out in the environment) to make the channel memoryless, resulting in a variation of the ALN channel: in case the molecule hits the receiver before its lifetime ends, we get $Z=X+T$, with $T$ following a truncated L\'evy distribution. By writing the expansion $I(X;X+T)=h(X+T)-h(T)$, the particular form of the L\'evy distribution is used to find a closed form expression of the entropy of its truncated version $h(T)$. The term $h(X+T)$ is bounded from below via the entropy power inequality (EPI), and from above by logarithm of the support of $X+T$.

\textbf{The discrete delay-selector model:} Let us consider the following discrete model of the timing channel . We divide the time horizon into time slots of duration $T_s$. At the beginning of each time slot, up to $\mathsf N$ indistinguishable molecules may be released. Molecules are not lost during transmission, and each of these molecules will eventually arrive at the receiver in the current time-slot or in the subsequent time slots. Furthermore, a channel delay of at most $\Delta\in\mathbb{N}$ is assumed: a transmission in time slot $j$ arrives in any of the following $\Delta$ time slots, \emph{i.e.,} in one of the time slots $j$, $j+1$, ..., $j+\Delta-1$. The receiver can count the number of molecules received in each time slot, but does not otherwise know the exact arrival times of individual molecules within a time slot. Overall, the input can be characterized a sequence $(x_1, x_2, \ldots, x_n)$ where $x_i\in [0:\mathsf N]$ denotes the number of molecules released in time slot $i$. The output is a sequence $(y_1, y_2, \ldots, y_n)$ where $y_i$ indicates the number of molecules received in time slot $i$. Since $Y_i$ depends not only on $X_i$, but also on $X_{i-1}, X_{i-2}, \cdots, X_{i-\Delta+1}$, this is a channel with memory.

The above model was originally introduced in \cite{Timing2} and called the \emph{delay-selector} model. Authors in \cite{Timing2} studied the normal Shannon capacity for the case of $\mathsf N=1$. The asymptotic behavior  of capacity in terms of the number of molecules and time intervals for communication is studied in \cite{EckfordChae}. Its zero-error capacity was completely characterized in \cite{Timing10} as $\log r$ where $r$ is the unique positive real root of the polynomial $x^{\Delta+1}-x^\Delta-\mathsf N=0$. The authors of \cite{Timing10} were not apparently aware of \cite{Timing2}, as \cite{Timing2} is not cited and their zero-error capacity result is not compared with the vanishing error result of \cite{Timing2}.

One of the main ideas used in \cite{Timing2} can be summarized as follows:\footnote{This is not the way the idea is presented in the paper itself.} consider a channel with memory. To compute the capacity, one would need to consider $n$-letter mutual information terms:
\cite{VerduHan}
$$\liminf_{n\rightarrow\infty} \frac{1}{n}\max_{p(x^n)}I(X^n;Y^n).$$
But $p(y^n|x^n)$ cannot be expressed as $\prod_{i=1}^n p(y_i|x_i)$. To derive a lower bound, let us take an i.i.d.\ input pmf $p(x^n)=\prod_{i=1}^n p(x_i)$.\footnote{This idea has been used in earlier works such as \cite{Shamai}.} Then, one gets a single-letter expansion as follows:
\begin{align*}I(X^n;Y^n)&=H(X^n)-H(X^n|Y^n)
\\&=\sum_{i=1}^n H(X_i) -H(X^n|Y^n)
\\&=\sum_{i=1}^n H(X_i) -H(X_i|Y^nX_{1:i-1})
\\&\geq \sum_{i=1}^n H(X_i) -H(X_i|Y_i)
\\&= \sum_{i=1}^n I(X_i;Y_i).
\end{align*}

Authors in \cite{Timing10} find an exact zero-error result for the delay-selector channel. They also provide explicit capacity achieving codes and a
linear-time decoding algorithm for their codes. To define a zero-error codebook, we need a few definitions: we write that $(x_1, x_2, \ldots, x_n)\leadsto (y_1, y_2, \ldots, y_n)$ if there is a way to obtain $(y_1, y_2, \ldots, y_n)$ from $(x_1, x_2, \ldots, x_n)$ on the delay-selector channel with appropriate choice of delays for individual molecules. A zero-error code consists of a class of input sequences such that for any two codewords $(x_1, x_2, \ldots, x_n)$ and $(x'_1, x'_2, \ldots, x'_n)$, one cannot find $(y_1, y_2, \ldots, y_n)$ such that
$(x_1, x_2, \ldots, x_n)\leadsto (y_1, y_2, \ldots, y_n)$ and $(x'_1, x'_2, \ldots, x'_n)\leadsto (y_1, y_2, \ldots, y_n)$ at the same time.

The paper follows by finding a recursive equation for the size of the optimal codebook of size $n$. A key observation in \cite{Timing10} is the following: if there are two codewords $\mathbf{x}=(x_1, x_2, \ldots, x_n)$ and $\mathbf{x'}=(x'_1, x'_2, \ldots, x'_n)$ such that $x_i=x'_i$ for $i>\Delta+1$, and $\sum_{i=1}^{\Delta+1}x_i=\sum_{i=1}^{\Delta+1}x'_i$, then $\mathbf{x}$ and $\mathbf{x'}$ may not both belong to a zero-error code at the same time. If this is the case and $q=\sum_{i=1}^{\Delta+1}x_i=\sum_{i=1}^{\Delta+1}x'_i$, then
one can obtain the output sequence $(0, 0, \ldots, 0, q, x_{\Delta+2}, x_{\Delta+3}, \ldots, x_{n})$ from both $\mathbf{x}$ and $\mathbf{x'}$.

\section{Capacity for the ligand-receptor model}\label{sec:ligand}
In this section, we review some results relating to the ligand-receptor models described in Section~\ref{sec:ligand-model-part}

\textbf{Memoryless binomial channel model:}
The ligand receptors are studied in \cite{EinSarFekITW11}, where a memoryless binomial distribution is proposed to model the number of bound receptors as the output while their binding probability is taken to be the input. Thus, the input to the channel is $p\in[0,1]$ and the output is a sample from $B(k,p)$ for some given $k$ receptors on the surface of the receiver. The distribution which maximizes the mutual information among this input-output is Jeffery's prior \cite{Xie}. Assuming an environmental noise, authors in \cite{AmiFarMirNasFek16} consider two bacterial point-to-point communication scenarios from the capacity viewpoint: (i) multi-type molecular communication with a single concentration level, and (ii) single-type molecular communication with multiple concentration levels. For both cases, upper and lower bounds are found on the capacity. For the upper bound, the symmetrized Kullback-Leibler divergence based upper bound of \eqref{KL} is employed. The lower bound is derived on the capacity under average and peak constraints by assuming a binary input. The approach of obtaining the lower bound is to covert the ligand-receptor model to a variation of Z-channel with the binary input and the outputs in $\{1, \ldots, N^\prime\}$, where $N^\prime$ is the total number of the receptors.

\textbf{Markov model:} In the Markov model with $k$ receptors on the surface of the receiver, the state of receiver is modeled by a vector in $\mathbb{S}=\{B, U\}^k$. Furthermore, the state  at time instance $i$ is  also the output of the receiver, $Y_{i}=S_i$. As pointed out in \cite{ThomasEckford} such state dependent channels belong to the class of channels studied (with and without feedback) in \cite{Chen05, Permuter14}. Even though we have a channel with memory, the main result is that  the capacity is achieved by i.i.d.\ input pmf $p(x^n)=\prod_{i}p(x_i)$ \cite{ThomasEckford} by directly investigating the analytical expression for mutual information. For the case of one receptor, output feedback is shown not to increase the capacity. Here, the important assumption is that when a receptor is in state $B$, its transition to state $U$ is independent of input. The intuition for this result is summarized in \cite{Ligand2} as follows: ``when we have a coding scheme which uses feedback, the encoding function depends on the output of the channel in the previous epochs. Since the channel has Markov structure, if we go back more that one epoch, we do not get useful information. Hence, one can modify the encoding function so that it would always assume that the previous output was $U$ (i.e., the receptor was at the unbound state). If the assumption was correct, it is similar to the feedback strategy. Otherwise, the state of the channel is $B$, and the next output is independent of the input. Thus, in both cases, the feedback strategy and the modified strategy have the same result. Therefore, every rate which is achievable via feedback can be achieved without feedback."  This argument only shows that feedback does not help when there is one receptor on the surface of the receiver. Authors  in \cite{EckfordThomasNew} consider the case of multiple ligand receptors. The resulting state dependent channels again would belong to the class of channels studied in \cite{Chen05}, and its feedback capacity can be computed via the formula given in \cite{Chen05}. However, explicit calculation of the capacity still requires solving an optimization problem. Authors in  \cite{EckfordThomasNew} work out this optimization problem and show that if one can vary the input concentration of molecules around the receiver arbitrarily fast, the capacity for $m$ ligand receptors can be found by simply multiplying $m$ times the capacity of a single receptor. Surprisingly, feedback does not help in this case either (\emph{i.e.,} revealing the current number of bound receptors to the transmitter cannot increase the communication rate).

\section{From point to point to multi-user:  the cascade problem}\label{sec:cascade}

Our focus up to this point has been on a point to point setting. But many of the envisioned applications of MC employ a network of molecular nodes. Molecular networks have received little attention in the literature. The problem is complicated by the fact that one needs to study the effect of \emph{memory} in the context of \emph{networks}; any realistic model of molecular medium or transceivers should consider the effect of {memory} of past actions in their formulation.

 Among different network structures, we focus on the cascade structure here in this section.
We provide an open problem
that was motivated by our study of molecular communication. As common with interdisciplinary topics, studying molecular communication may lead one to formulate new information theoretic problems.

\subsection{The cascade problem}
\begin{figure}
\begin{center}
\includegraphics[width=1\textwidth]{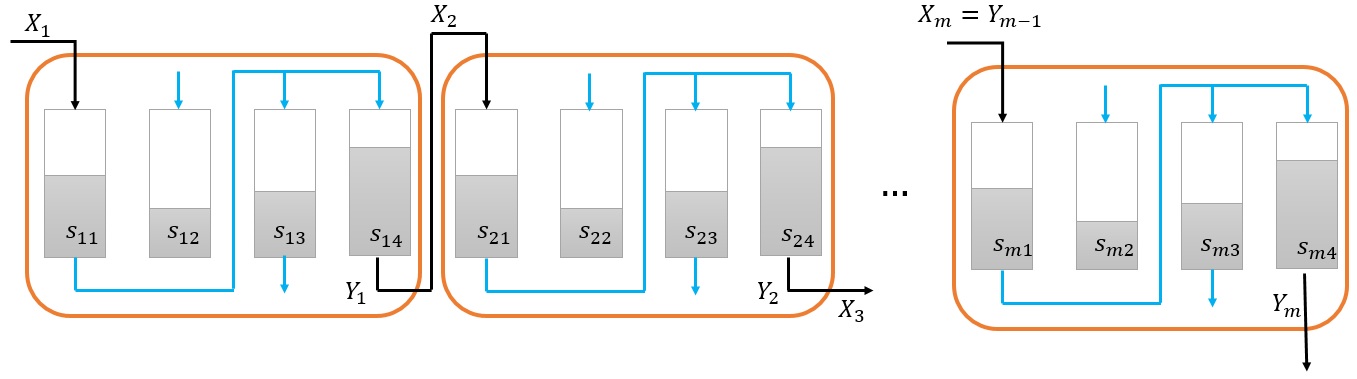}
\end{center}
\caption{Diagram of a chain of $m$ bacteria, with the output of each directly connected as the input of the next bacterium. Each bacterium has a state space determined by variables $s_1, s_2, s_3, s_4$.}\label{BacteriaCableFig}
\end{figure}

The channel cascade problem arises naturally when communicating over a medium consisting of identical objects placed one after another. For instance, this chain could consist
of biological cells as in the bacteria cable of \cite{Ubli}, where the output of each cell is directly
connected to the input of the next cell. There is no intelligent processor after each cells to decode and then encode the
message for the next cell. The diagram for the bacteria cable is depicted in Fig.~\ref{BacteriaCableFig}.   Each bacterium is modeled by a number of electron queues that describe its electron transport chain (a process completed by a cell to produce
energy). The size of these queues statistically affects transitions of electrons from its input terminal (electron donor) to its output terminal (electron acceptor). Here a vector of size  four, consisting of the length of the queues associated to each cell, represents the state variable associated to each cell. By placing the cells one after another, we obtain the cascade architecture. In this section, we provide a conjecture about cascade of channels with memory. This conjecture implies that the capacity of the cascade link goes to zero as the length of the cable goes to infinity.

\begin{figure}
\begin{center}
\includegraphics[width=0.95\textwidth]{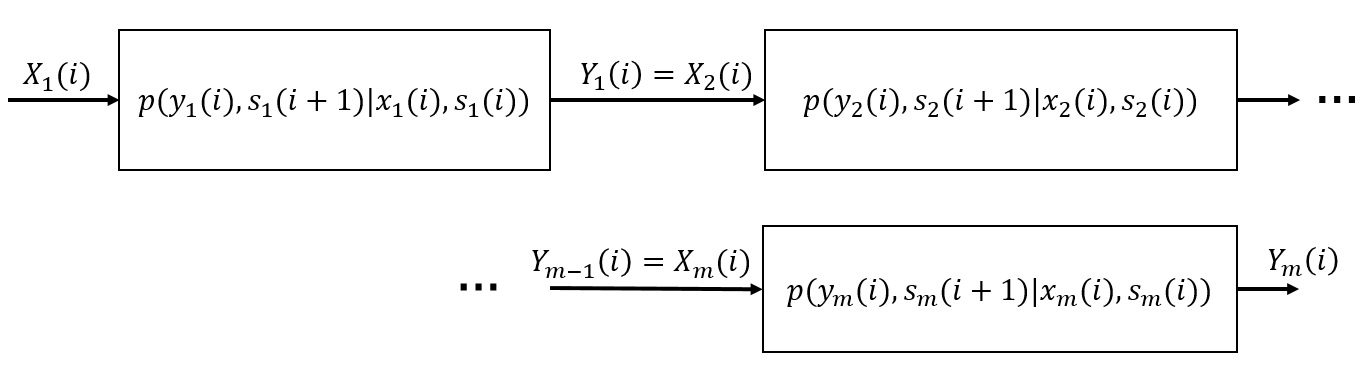}
\end{center}
\caption{Cascade of $m$ general channels with memory}\label{CascadeOfChannelMemory}
\end{figure}

Another motivation for the cascade problem is due to the short range of molecular communication and the need for multi-hop communication to send messages across longer distances.

\begin{problem}[Cascade Problem] Consider a state-dependent channel
\begin{align}
p(y(i), s({i+1})|x(i),s(i))\label{eqn:state_channel}
\end{align}
with identical input and output alphabets, \emph{i.e.,}  $\mathcal X=\mathcal Y$. Consider $m$ identical copies of this channel and let us denote the input, output and state of the $j$-th channel at time $i$ by $X_{j}(i)$, $Y_{j}(i)$ and $S_j(i)$ (see Fig.~\ref{CascadeOfChannelMemory}). We say that these channels are connected in cascade if $X_j(i)=Y_{j-1}(i)$, \emph{i.e.,} the output of the $(j-1)$-th channel is the input to the $j$-th channel (no intelligent processing unit is placed between the output of one hop to the input of the next hop). The input of the cascade channel is $X_1(i)$ and its output is $Y_m(i)$. We allow for block codes. A communication rate is said to be achievable if the average error probably converges to zero as the blocklength goes to infinity, \underline{regardless of the initial state of the $m$ channels}. The question is what is the behavior  of the capacity of cascade of $m$ replicas of this channel as a function of $m$?
\end{problem}

To best of our knowledge, cascade of channels with memory has not received much attention in the literature. The only result that we are aware of is that of \cite{Kramer} wherein authors consider cascade of a certain nonlinear channel with memory that appears in the context of optical fibers, but the analysis is specialized  to the particular optical channel that is governed by a certain stochastic differential equation. Capacity of a finite-state discrete channel with input $X$, output $Y$ and state variable $S$   is given by \cite[p.100]{Gallager}
\begin{align}\lim_{n\rightarrow\infty} \frac{1}{n}\max_{p(x^n)}\min_{s_0}I(X^n;Y^n|s_0).\label{eqn:narb}\end{align}
Since cascade of a channel with memory is itself a channel with memory, in principle it is possible to write a formula for channel capacity by considering (multi-letter) mutual information between the input of the first channel and the output of the $m$-th channel in cascade. The challenge is to study the limit as blocklength $n$ tends to infinity.\footnote{If we restrict to $n=1$ in the expression of equation \eqref{eqn:narb}, the result of \cite[p. 527]{Gallager} can be applied. When considering  single-letter mutual information between the input of the first channel and output of the last channel, each of the $m$ finite-state channels has some initial state and thereby a channel transition matrix. Thus, the end-to-end mutual information can be found using the result of \cite[p. 527]{Gallager} for memoryless channel corresponding to these transition matrices.}

To state our main conjecture about the cascade problem, we need a definition:
\begin{definition}
We define the concept of zero-error capacity $C_0$ for a channel with memory of \eqref{eqn:state_channel}, as the maximum rate of information that can be communicated with \emph{exactly} zero error, regardless of the initial state, $s(0)$, of the channel.\footnote{This corresponds to the zero-error version of $\underline{C}$ as in \cite[p.100]{Gallager}.} The zero-error capacity is defined when we allow for multiple-use of the channel with memory (blockcoding is allowed), but only codes  with exactly zero error (and not a vanishing error probability as blocklength tends to infinity) are allowed.

Given a channel with memory, observe that its $m$ cascade channel is itself a channel with memory (with its state being the state vector of the $m$ channels), and hence its zero-error capacity can be defined in the same way. Let $C^m_0$ denote the zero-error capacity of $m$ cascade channels. Also, let $C^{\inf}_0=\lim_{m\rightarrow\infty}C^m_0$.
\end{definition}

We can now formally state our conjecture:
\begin{conjecture}
Take an arbitrary state dependent channel $p(y(i), s({i+1})|x(i),s(i))$ with finite input/output alphabet $|\mathcal X|=|\mathcal Y|<\infty$.  Then, the Shannon capacity of the $m$ cascade channel converges to $C^{\inf}_0$ as $m$ goes to infinity if $|\mathcal{S}|<\infty$, \emph{i.e.,} if it is a finite state channel.  On the other hand, one can find channels with finite input/output alphabet but with infinite memory, $|\mathcal{S}|=\infty$, such that the Shannon capacity of the cascade channel converges to a value that is strictly larger than $C^{\inf}_0$ as $m$ goes to infinity.
\end{conjecture}

A partial proof of the above conjecture is given in Appendix \ref{secprofconj}.

\subsection{Cascade of memoryless channels}
Memoryless channels are special cases of channels with memory, and their study is the first step towards solving the general problem with memory. Consider a \emph{memoryless} channel $p_{Y|X}$ with no state variable. We have the following Markov chain:
$$X_1\rightarrow Y_1=X_2\rightarrow Y_2=X_3\rightarrow\cdots\rightarrow Y_{m-1}=X_m\rightarrow Y_m,$$
where $p_{Y|X}(y_i|x_i)$ is given. If we denote the transition matrix for channel $p_{Y|X}$ by $\mathsf P$, the transition matrix for the $m$ cascade channel will be equal to $\mathsf P^m$. Then, for a memoryless channel we have  $C^{m}_0(\mathsf{P})=C_0(\mathsf P^m)$.

It is insightful to construct a Markov chain as follows: let the state space be $\mathcal{X}=\mathcal{Y}$ and the transition probability from state $x$ to state $x'$ be equal to $p_{Y|X}(x'|x)$. Then, random variables $X_1, X_2, \cdots, X_m, Y_m$ can be understood as a random walk sequence on this Markov chain with the initial state of $X_1$.

\textbf{Memoryless channels with finite input/output alphabets:}
As an example, cascade of $m$ Binary Symmetric Channels (BSCs) with parameter $p$ is a BSC channel with parameter $q=(1-(1-2p)^m)/2$, and capacity $1-h(q)$ where $h(\cdot)$ is the binary entropy function. By writing the Taylor expansion of entropy function around $1/2$, it is easy to verify that the capacity drops exponentially fast in $m$. There are few previous works on the behavior of memoryless cascade channels in the literature, most of which consider cascade of simple channels  (\emph{e.g.} see
\cite[p. 527]{Gallager},  \cite[p.85]{cas__ref1}, \cite[p. 221]{cas__ref2}, \cite[p. 42]{cas__ref3}, \cite{cascade_1, Majani}, \cite{Simon, Silverman}). As a historical note, just like Shannon who was motivated by the repeatered telephone lines common in his time, the motivation of Simon for studying the cascade channel was the repeatered telephone line  \cite{Simon}.

It turns out that  just like the BSC example, for any memoryless channel with finite alphabet and positive transition matrix, \emph{i.e.,} $p(y|x)>0$ for all $x,y$, the Shannon capacity of the cascade channel drops exponentially to zero. In fact, to compute the capacity, we seek the mutual information between the initial state $X_1$ and the final state $Y_m$. If the Markov chain has a finite state space, and it is irreducible and aperiodic (implied by $p(y|x)>0$ for all $x,y$), it will have a unique stationary distribution to which the chain converges (exponentially fast), starting from any initial state. Convergence of the Markov chain to a stationary distribution implies that the correlation between initial state $X_1$ and the final state $Y_m$ after $m$ walks fades away exponentially fast. This would imply that the capacity drops to zero exponentially fast. To see why the drop occurs exponentially fast in another way, note that by the data processing inequality, we have
$$I(X_1;Y_m)\leq I(X_{2};Y_m)\leq \cdots \leq I(X_{m};Y_m).$$
However, the above data processing inequality can be strengthened by finding a constant $\eta<1$ such that for any $U\rightarrow X\rightarrow Y$ we have $I(U;Y)\leq \eta\cdot I(U;X)$, instead of the weaker $I(U;Y)\leq I(U;X)$ \cite{AG, AGKN13}. We can then write that
$$I(X_1;Y_m)\leq \eta I(X_1;Y_{m-1})\leq \cdots \leq \eta^{m-1} I(X_{1};Y_1)\leq \eta^{m-1} H(X_{1}),$$
showing that $(X_1;Y_m)$ drops to zero exponentially fast in $m$.

Let us consider a general discrete memoryless channel $p(y|x)$. Let us first consider the zero-error capacity of the cascade channel, \emph{i.e.,}  the rate at which it is possible to send information with \emph{exactly} zero error. First, we provide some definitions from finite-state Markov chain theory. State $j$ is said to be accessible from state $i$ (shown by $i\rightarrow j$), if it is possible to reach state $j$ from state $i$ in some number of steps, \emph{i.e.,}  if there exists some $n\geq 0$ such that $p^{(n)}_{ij}>0$, where $[p_{ij}]$ is the probability transition matrix. Two distinct states $i$ and $j$ are said to communicate (shown by $i\leftrightarrow j$), if state $j$ is accessible from state $i$ and  state $i$ is accessible from state $j$. A state is always considered  to communicate with itself.
We can partition the states of a Markov chain into disjoint communicating classes, where two states $i$ and $j$ are in the same class if and only if $i$ and $j$ communicate ($i\leftrightarrow j$). A communicating class is closed if starting from a state in that class, we will remain in the class forever (i.e., states outside the class are not accessible from the states that belong to this class). A state is called \emph{recurrent} if starting from the state, the probability of returning to it is one. Otherwise, it is called \emph{transient}.  The states of a closed communicating class are all {recurrent}, and the states of a non-closed communicating class are all {transient}.

With the Markov chain interpretation of the channel in mind, the zero-error capacity of the cascade of $m$-channels will be positive if the chain has more than one closed communicating class. The reason is that if initial state belongs to one of the closed communicating classes, it will remain in that class forever. Therefore, it is possible to use the identity of the communicating class for signaling. Furthermore, observe that zero-error communication at a positive rate is possible even when the chain is irreducible and has only one closed communicating class, if the states are periodic with period $T$. In this case, we can partition the graph into $T$ components and index them by $0, 1, \cdots, T-1$. Then, starting from an initial state in component $j$, after passing through $m$ cascade channels, we will end up in state $(j+m)\mod T$. In fact, one can show the following:
\begin{proposition}\label{proposition1} Let $C_0(\mathsf P^m)$ and $C(\mathsf P^m)$ denote the zero-error and Shannon capacities of the cascade of $m$ repetitions of a finite alphabet memoryless channel with transition matrix $\mathsf{P}$. Then, $$\lim_{m\rightarrow\infty} C(\mathsf P^m)=\lim_{m\rightarrow\infty} C_0(\mathsf P^m)=\log(\sum_{i=1}^r T_i)$$ where $r$ is the number of closed communicating classes of the Markov chain corresponding to $\mathsf{P}$, and $T_i$ is the period of nodes in the $i$-th class (thus, input alphabet $\mathcal{X}$ can be partitioned into $r$ closed communicating classes, and set of transient nodes).
\end{proposition}
The proof is given in Appendix \ref{appndixA}.

\textbf{Memoryless channels with infinite input/output alphabets:}
For memoryless channels with countably infinite or {continuous} alphabets, the behavior  of the capacity of cascade channels can be very different than the finite discrete channels. To see why the behavior  may be different, note that if we cascade $m$ memoryless Gaussian channels with noise $\sigma^2$, the overall channel will be a Gaussian channel with noise variance $m\sigma^2$. With an input power constraint $P$, the capacity of the cascade channel will be $\frac{1}{2}\log(1+P/(m\sigma^2))$ which is equal to $P/(2m\sigma^2)$ for large values of $m$. Thus, the capacity is proportional to $m^{-1}$, and not exponentially decreasing in $m$. The difference between the infinite alphabet and finite alphabet cases is that while it is still possible to construct a Markov chain on the state space $\mathcal{X}=\mathcal{Y}$, the chains on infinite alphabet spaces can be more involved and may not even have a stationary distribution.

 As far as we are aware, the possible behaviors  of the capacity of cascade of identical memoryless channels with countably infinite or continuous alphabets have not received any attention. More specifically, one may ask that besides exponential and $m^{-1}$ drop in capacity, what other behaviors  are possible? An example of a memoryless channel with countably infinite input/output alphabet is provided below to show that unlike the finite alphabet case, one can have
$$\lim_{m\rightarrow\infty} C(\mathsf P^m)> \lim_{m\rightarrow\infty} C_0(\mathsf P^m).$$

\begin{figure}
\begin{center}
\begin{tikzpicture}[ ->,>=stealth',shorten >=1pt,auto,node distance=3.3cm,
                thick,main node/.style={circle,draw,font=\Large\bfseries}]
  \node[main node] (1) {0};
  \node[main node] (2) [right of=1] {1};
  \node[main node] (3) [right of=2] {2};
  \node[main node] (4) [right of=3] {3};
  \path
    (1) edge [loop above] node {1} (1)
    (2) edge [bend left]  node [above] {$1-a_1$} (1)
	edge node [above] {$a_1$}(3)
    (3) edge [bend left] node [above] {$1-a_2$} (1)
	edge node[above]{$a_2$}(4)
    (4) edge [bend left] node[below] {$1-a_3$}(1);

\draw [ultra thick, fill] (12.5,0) circle [radius=0.025];
\draw [ultra thick, fill] (12,0) circle [radius=0.025];
\draw [ultra thick, fill] (12.25,0) circle [radius=0.025];
 \draw[->, thick] (10.3,0) to (11.5,0);
\end{tikzpicture}
\end{center}
\caption{The Markov chain of Example \ref{exmple1}, defining a channel on $\mathcal{X}=\mathcal{Y}=\{0,1,2,\cdots\}$.}\label{figMarkov}
\end{figure}
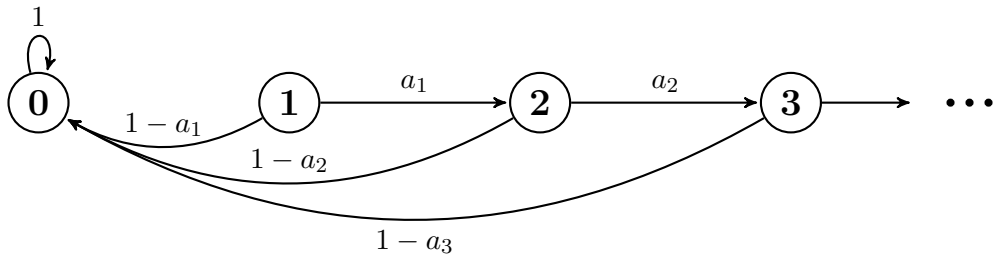

\begin{example}\label{exmple1}
Let $\mathcal{X}=\mathcal{Y}=\{0,1,2,\cdots\}$ be the input/output alphabets of a memoryless channel with the following transition probabilities: $p_{Y|X}(i+1|i)=a_i$ and $p_{Y|X}(0|i)=1-a_i$. This is depicted in Fig.~\ref{figMarkov}.  In other words, being at state $i$, with probability $a_i$ we go to state $i+1$ and with probability $1-a_i$ we go to state $0$. Assume that $a_0=0$, \emph{i.e.,}  we always stay in state $0$ if we end up there.
The probabilities $a_i$ are chosen such that $b_m=\prod_{i=1}^m a_i$ for $n\geq 1$ is a decreasing positive sequence that converges to $1/2$ as $m$ tends to infinity. Observe that zero-error capacity $C_0=0$ for this channel because given any input, output symbol $0$ occurs with some positive probability. Therefore, the zero-error capacity of the cascade channel is also zero.

Let us assume the uniform input distribution on $\{0,1\}$ (input power is $1/2$). Then, if we use input $X_1=0$ on the cascade of $m$ identical channels as above, the output will be $Y_m=0$ because input $0$ is always mapped to output $0$. For input $X_1=1$, with probability $b_m$, the output at the cascade of $m$ channels will be equal to $m$; with probability $1-b_m$, it will be equal to $0$. Therefore, the cascade of $m$ channels is essentially a $Z$-channel with input alphabet $\{0,1\}$ and output alphabet $\{0,m\}$. The input/output mutual information as $m$ tends to infinity, converges to the input/output mutual information of a $Z$-channel with parameter $1/2$, which is non-zero.
\end{example}

One implication of the above example is as follows: given an integer $L$, let us consider the ``$L$-truncated version" of the above channel as follows: the input/output state space are $\{0,1,2,\cdots, L\}$, and $p_{Y|X}(i+1|i)=a_i$ and $p_{Y|X}(0|i)=1-a_i$ for $0\leq i<L$. For $i=L$, $p_{Y|X}(0|L)=1-a_L$ and $p_{Y|X}(j|L)$ for $1\leq j\leq L$ is defined arbitrarily.  The above example shows that the capacity of the ``$L$-truncated version" of the above channel converges to zero as $m$ tends to infinity \emph{for any arbitrarily large $L$}, but the capacity of the cascade of the channel itself is positive as $m$ tends to infinity, even in the presence of an input power constraint.

Despite the above negative result, cascade capacities of many continuous alphabet memoryless channels (with $C_0=0$) converge to $0$ as the length of the cascade channel goes to infinity. This can happen if  the strong data processing constant ($\eta$) is less than one. Even when $\eta=1$, the idea of ``non-linear" data processing constant of Polyanskiy and Wu \cite{Polyanskiy} may be still used to show that mutual information drops to zero. The idea is to show an appropriate (possibly non-linear) increasing function $f(\cdot)$ satisfying $f(t)\leq t$, such that for any $U\rightarrow X\rightarrow Y$ we have $I(U;Y)\leq f(I(X;Y))$. This would then imply that
\begin{align*}I(X_1;Y_m)&\leq f(I(X_1;Y_{m-1}))
\\&\leq (f\circ f)(I(X_1;Y_{m-2}))
\\&\leq\cdots
\\&\leq (f\circ f\cdots \circ f)(I(X_{1};Y_1)).\end{align*}
Therefore, one needs to look at the convergence of the sequence of $a, f(a), f(f(a)), \cdots$ for $a=I(X_1;Y_1)$.

\textbf{Variation of the problem with relay nodes:}
A natural variation of the cascade problem with relay processing nodes placed in between any two consecutive memoryless channels is studied in \cite{Polyanskiy, Line, svl13, sub12}.  If there is no restriction on the relays, they can decode and re-encode the  information. In this case, the Shannon capacity of the cascade channel will be equal to the min-cut capacity of the links. Furthermore, the zero-error capacity of the cascade channel will be equal to $C_0(\mathsf P)$, the zero-error capacity of each individual channel. Therefore, there are potential improvements both in terms of the Shannon capacity and the zero-error capacity.

To approach the min-cut capacity, one needs to use large blocklengths. An interesting result for the cascade problem with  relays is provided in \cite{Line}.  It is shown in \cite{Line} that if the relays are forced to process blocklength of fixed size, then for any discrete memoryless channel, the cascade capacity converges to the zero-error capacity $C_0(\mathsf P)$
 exponentially fast as the length of the cascade channel $m$ goes to infinity. In other words, relays of limited complexity are not helpful.

Another interesting result for additive Gaussian channel is given  in  \cite{Polyanskiy}. It considers the simple strategy of each relay comparing the input signal with a threshold and choosing the input to the next channel accordingly. By a judicious choice of the thresholds, it is shown that the
end to end mutual information drops like ${\log\log m}/{\log m}$ (rather than $m^{-1}$), but the thresholds used depend on the location of the relay in the cascade network.

\subsection{Channel with memory}
The channel memory allows the channel statistic $p(y_i|x_i,s_i)$ to adapt itself according to the channel state $s_i$ (itself influenced by previous inputs to the channel). According to our conjecture, this freedom is limited and essentially useless for finite state channels. Observe that even in a finite state channel ($|\mathcal S|<\infty$), the output $Y_i$ may still depend on all previous inputs $X_i, X_{i-1}, X_{i-2}, \cdots$. This is because $Y_i$ depends on $X_i$ and $S_i$, but $S_i$ may be affected by the \emph{entire} past inputs. Since the input $X_i$ is not affected by only a finite number of past inputs, one cannot try to model the memory effect with the actions of the relays in the model studied in \cite{Line}, wherein relays are forced to process blocklength of fixed size.

\section{Concluding remarks and some open problems}\label{sec:concl}

 Efficient communication among small devices is an area where communication and information theorists can contribute. In many applications, communication among nano-devices is not a goal in itself, but occurs with the goal of serving a task. In other words, there is not always a given explicit messages of a certain rate that needs to be communicated; but the message itself is a parameter that needs to be created and transmitted. Examples of such scenarios from classical information theory are studies of the relation of communication and control, communication for coordination or function computation. These may need rethinking to address the specific setting of applications in molecular communication. For instance, proper mathematical models of the restrictions imposed by the transmitter and receiver may be needed; finite blocklength and one-shot results may be of importance; and channel memory may require serious considerations. As discussed in the paper, study of networks with memory is challenging, even for the simple cascade architecture. Next, results from arbitrarily varying channels in classical information theory might be of particular relevance because of  stochastic effects such as sensitivity of the medium to temperature or jitters. Finally, extremely long delay of the diffusion process complicates formation of feedback links, which are both useful for effective communication and providing stability in control.

There are already many ongoing developments in the literature. Many communication oriented papers focus on error probability and consider simple ISI-mitigation techniques. To put these results in a firmer theoretical footing, it would be interesting if one can show near optimality of these techniques from an information theoretic perspective (if one can restrict to energy efficient coding strategies of limited complexity). We introduced three models for the transmitter and four models for receiver. While ligand receiver is a realistic receiver model, it has not been studied in conjunction with any of the transmitter models. In particular, it would be interesting to characterize the capacity of the ligand receptor with the Poisson concentration transmitter. Also, further works on mathematical modeling of different components of a molecular communications system (transmitter, receiver and channel) and their intrinsic noises and temporal variations are needed for any thorough information theoretic analysis. For instance, a transmitter may not be able to \emph{instantaneously} release molecules in the environment \cite{ArjAhmSchKen16}.

As mentioned in the introduction, development of a proper theoretical framework for studying limitation of resources in the context of MC is necessary, perhaps in the context of specific circuit models. For instance, for the VLSI model, in a series of works, \cite{Grover1, Grover2} consider how much information bits need to travel across the surface of a VLSI circuit in order to implement an encoder or decoder function. The concept of ``frictional losses" associated with moving information on a substrate is developed for Thompsons VLSI-model and used to characterize fundamental energy requirements on encoding and decoding in communication circuitry. It may be possible to develop proper models for molecular circuits.
 Molecular circuits, including logic gates or processing units, could be implemented using chemical reactions. It is not clear how the notion of complexity in the context of MC should be defined at this point. It could be the ATP consumption at the encoder and decoder units \cite{energyLu}, the number of reaction (in chemical computation) or the length of the DNA sequences used (in DNA computation). See also \cite{EnergyModelFarsad}.

Finally, it is also worthwhile to look for new signalling mechanisms. To the best of our knowledge, the common presumption in the existing literature is that the information should be coded by the transmitter via the action of \underline{releasing molecules} in the environment. However, other options are possible too. For instance, we might have a node (separate from the transmitter) that emits molecules in the environment according to some predefined deterministic pattern. The transmitter may change the communication medium, by exploiting chemotaxis or changing the flow (or in general physical properties) of the medium between the emitter and receiver. We may call this ``molecular media based modulation" as it resembles media based modulation proposed in the classical communication \cite{Media-Comm}. The closest existing work to molecular media based modulation appears to be \cite{NakanoSuda}. Further understanding of molecular media based signaling   can be of interest.

\section*{Acknowledgement}
The authors would like to thank Prof. Urbashi Mitra for bringing our attention to  bacteria cables, and thank Hamidreza Arjmandi, Reza Mosayebi, Gholamali Aminian, Ladan Khaloopour and Mehdi Soleimanifar for helping with figures and some data. We also like to thank the anonymous reviewers for their comments that improved the presentation of this paper.

\appendix
\section{Partial proof of the conjecture}\label{secprofconj}
\begin{proof}[Partial proof of the conjecture] We construct an example for the second part of the conjecture by carefully constructing a channel with memory. A channel with infinite memory can remember all of its past inputs and can adapt its input-output statistical description according to all past inputs. The general idea is to design the state dependent channel with memory, $p(y(i), s({i+1})|x(i),s(i))$, that behaves intelligently as a memoryless channel followed by a relay node who can decode and re-encode the information. Therefore the cascade of the $m$ channels with memory behaves as the cascade of $m$ memoryless channels with relay processers placed between them. These relays decode and re-encode information allowing for the capacity of the cascade channel to reach the min-cut capacity of the links, strictly above $C_0^{\text{inf}}$.

We construct a $p(y(i), s({i+1})|x(i),s(i))$ structured as follows and depicted in Fig. \ref{Cascade2n2}:
The input is $X=(X_1, X_2)$ and the output is $Y=(Y_1, Y_2)$, where $X_1, Y_1\in\{0,1\}$, and $X_2, Y_2\in\{1,2,\ldots, k\}$. We assume that $Y_1=X_1$ regardless of the state value, and input $X_2$. The input $X_2$ first passes through a memoryless erasure channel $p(x'_2|x_2)$, with erasure probability $e$ acting independent of the state, and then through a state-dependent channel $p(y_2|x'_2, s)$. The value of $s$ is updated according to the input $(x_1, x'_2)$ and its previous value. The idea is to have $p(y_2|x'_2, s)$ acting as a relay node that decodes information over the erasure channel and re-encodes it.

\begin{figure}
\begin{center}
\includegraphics[width=0.8\textwidth]{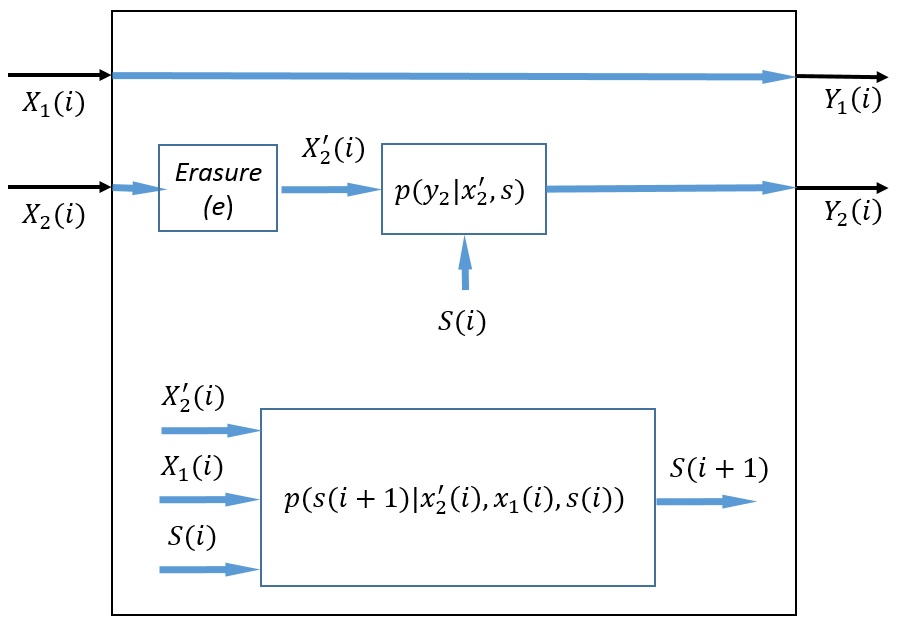}
\end{center}
\caption{Description of a single channel with memory used in partial proof of the conjecture. Here $Y_1=X_1$ are binary, while $X_2$ and $Y_2$ are k-ary. The alphabet for $X'_2$ is $\{e, 1,2,\cdots, k\}$ is created by passing $X_2$ through a memoryless erasure channel.}\label{Cascade2n2}
\end{figure}

\begin{figure}
\begin{center}
\includegraphics[width=1\textwidth]{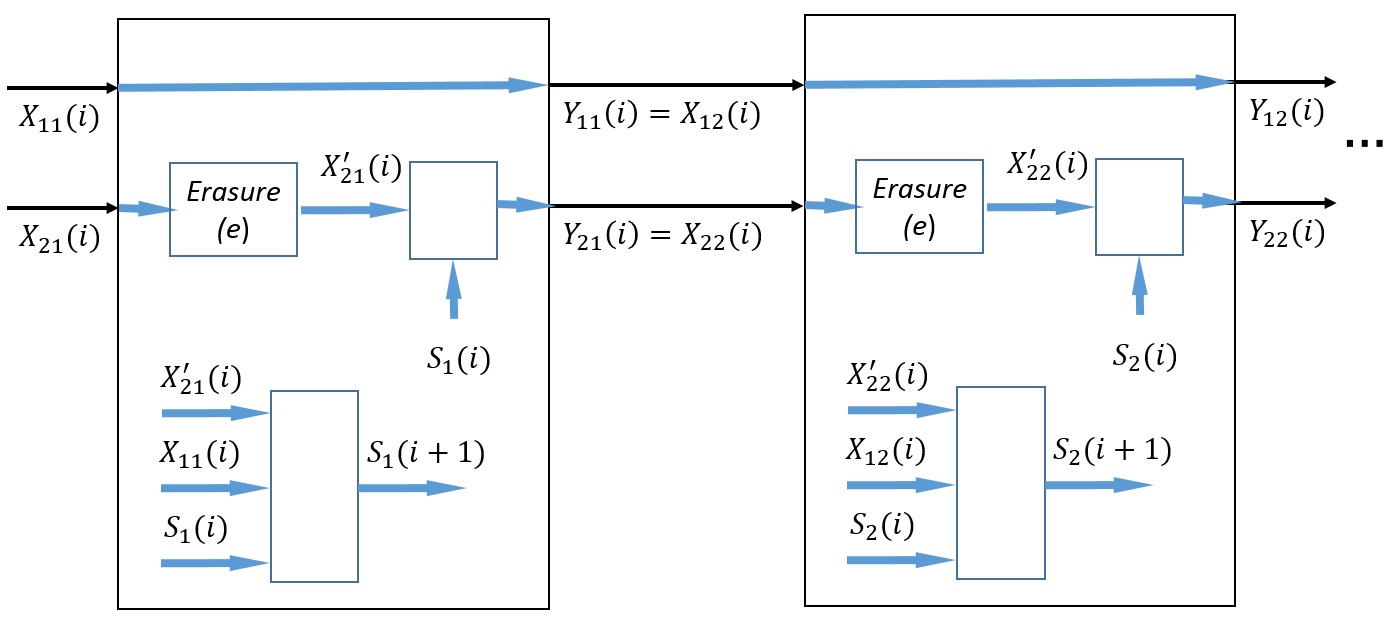}
\end{center}
\caption{Diagram of a chain of channels connected to each other in series. Observe that the $X_1$ part passes through the chain without being disturbed.}\label{Cascade2n}
\end{figure}

Note that $C_0^{\text{inf}}=1$ for this channel. Due to the $X_1$ part of the input that is directly connected to the output $Y_1$, it is possible to send one bit throughout the $m$ cascade channels (see Fig.~\ref{Cascade2n}). Furthermore, because an erasure channel is acting on the $X_2$ part right after entering the first block, the input part $X_2$ can be erased with some positive probability. Therefore, the $X_2$ part cannot be used to transmit information with exactly zero error probability.

Observe that the $X_1$ component passes through all the cascade nodes without any error. It is used for the following actions:
\begin{itemize}
\item To reset the initial state of all of the state-dependent channels in the cascade structure to a known initial ``reset" state. We assume that this is done by sending three consecutive zeros ($000$) on the $X_1$ part.
\item Once in the initial ``reset" state, and given the error probability $\epsilon$ at the transmitter, the transmitter chooses a blocklength $N$, finds the binary expansion of $N$ and send its bits to each of the state-dependent channels on the $X_1$ part. This incurs $\log(N)$ bits to communicate. The end of the $\log(N)$ bits is marked by the ``end sequence" string $001$. In order to avoid confusion with ``reset" and ``end sequence" strings $000$ and $001$, we assume that blocklength $N$ is chosen such that its binary expansion does not have either of $000$ and $001$ showing up as its consecutive digits. Thus far, the channel has been used $3+\log(N)+3$ times. 
\item Once the blocklength $N$ is conveyed, the $X_1$ link is used to communicate information  $N$ more times, while ensuring that the reset symbol $000$ is not used accidentally. One way to achieve this is to code the information bits into sequences of bits that do not have any two consecutive zeros. The number of such sequences of length $N$ is given by the Fibonacci number \cite[p. 253]{Fibo}, growing like $((1+\sqrt 5)/2)^N$. Thus, with this set of sequences of length $N$, we may convey $N\log(1+\sqrt 5)/2$ bits of information. In other words, the communication rate is $\log(1+\sqrt 5)/2$, which is less than $1$ bit per symbol that we could have achieved on the $x_1$ with simple transmission of bits; we will compensate for this hit using the $x_2$ input part.
\end{itemize}

The state-dependent channel can remember all the past inputs. Once it goes to a reset state, it starts learning the blocklength $N$. Once that is over, each node knows exactly the blocklength $N$; then, all of the nodes can agree on an appropriate decoder/encoder of blocklength $N$ for communication over an erasure channel with paramter $e$. Each node begins looking at the $X_2$ part of their input for the next $N$ symbols. To produce $Y_2$, each node simulates the appropriate decoder/encoder for the blocklength $N$ on the erasure channel. Because the capacity of the erasure channel is $(1-e)\log(k)$, this would allow for reliable transmission of $N((1-e)\log(k)-\epsilon)$ bits of information with error probability $\epsilon$ on each hop. The total error probability will be at most $m\epsilon$ where $m$ is the length of the cascade. This can be made arbitrarily close to zero for a given $m$ by sending blocklength $N$ to infinity and $\epsilon$ to zero.

The total number of bits that are communicated is $N\log(1+\sqrt{5})/2+N((1-e)\log(k)-\epsilon)$, and the channel is used $3+\log(N)+3+N$ times. This gives us the rate $\log(1+\sqrt 5)/2)+(1-e)\log(k)$ by letting $N$ converge to infinity and $\epsilon$ converge to zero. Observe that $\log(1+\sqrt 5)/2+(1-e)\log(k)$ can be made arbitrarily larger than $C_0^{\text{inf}}=1$ by making $k$ large.
\end{proof}

\section{Proof of Proposition \ref{proposition1}}\label{appndixA}
Observe that $C_0(\mathsf P^m)\geq \log(\sum_{i=1}^r T_i)$ for any $m$. To see this, note that by choosing an arbitrary vertex from  communicating class $i$ and partition $j\in\{0,1,2,\cdots, T_i-1\}$, we can perfectly predict that after going through $m$ channels, the final state is in communicating class $i$ and partition $k=(m+j)\mod T_i$. Therefore, we can use our choice of communicating class and one of its partitions for signaling with zero error probability. Thus,
$$\lim_{m\rightarrow\infty} C_0(\mathsf P^m)\geq \log(\sum_{i=1}^r T_i).$$
Because for each $m$, $C(\mathsf P^m)\geq C_0(\mathsf P^m)$, it only remains to show that
$$\lim_{m\rightarrow\infty} C(\mathsf P^m)\leq \log(\sum_{i=1}^r T_i).$$

Let us assume that $\mathcal X=\bigcup_{i=0}^r \mathcal{V}_i$ where $\mathcal{V}_0$ is the set of transient nodes, and $\mathcal{V}_i$ for $i\in [1:r]$ is the $i$-th closed communicating class. Furthermore, for $i\in[1:r]$, the induced graph on  $\mathcal{V}_i$ is $T_i$-partite and we can correspondingly partition $\mathcal{V}_i$ into $T_i$ sets $\mathcal{V}_{ij}$, having $\mathcal{V}_i=\bigcup_{j=1}^{T_i}\mathcal{V}_{ij}$.

It suffices to show that for any arbitrary $p(x_1)$ on $X_1$,
$$\lim_{m\rightarrow\infty} I(X_1;Y_m)\leq \log(\sum_{i=1}^r T_i).$$
We prove this statement in three steps.

\emph{Case (i): the support of $p(x_1)$ is a subset of $\mathcal V_{ij}$ for some $i\in[1:r], j\in[1:T_i]$}. In other words, $X_1$ is in the $j$-th partition of the $i$-th communicating class. Observe that $I(X_1;Y_m)$ is a decreasing sequence in $m$, by the data processing inequality. Therefore, it suffices to study the limit for the subsequence defined by $m=\ell T_i$ for $\ell=1,2,\cdots$. But if $X_1$ is in the $j$-th partition of the $i$-th communicating class, after $\ell T_i$ steps, it will return to the same partition of the same communicating class. Therefore, we can define a reduced Markov chain on nodes in the $j$-th partition of the $i$-th communicating class that specifies the transition probabilities after $T_i$ steps. This Markov chain is irreducible and aperiodic. Therefore, it has a unique stationary pmf $\pi$ to which the chain converges regardless of initial state $X_1$. Thus, $H(Y_{\ell T_i})$ and $H(Y_{\ell T_i}|X_1=x_1)$ both tend to the entropy of $\pi$ as $\ell$ converges to infinity, for any arbitrary $x_1$ in $\mathcal V_{ij}$. This implies that $\lim_{m\rightarrow\infty} I(X_1;Y_m)$ is zero in this case.

\emph{Case (ii): $p(x_1)=0$ for all $x_1\in\mathcal V_0$.} Let random variable $Q\in\{(i,j): i\in [1:r], j\in[1:T_i]\}$ denote the index of the communicating class and the corresponding partition that $X_1$ belongs to. From our earlier discussion $Q$ is a deterministic function of both $X_1$ and $Y_m$. Thus, \begin{align*}I(X_1;Y_m)&=H(Q)+I(X_1;Y_m|Q)\\&\leq
\log|\mathcal Q|+I(X_1;Y_m|Q)
\\&=\log(\sum_{i=1}^r T_i)+\sum_{q}I(X_1;Y_m|Q=q)p(Q=q).
\end{align*}
By case (i) and the fact that $p(x_1|q)$ is concentrated on one of the partitions of a communicating classes, we have that for any $q$:
$$\lim_{m\rightarrow\infty} I(X_1;Y_m|Q=q)=0.$$
Therefore,
$$\lim_{m\rightarrow\infty} I(X_1;Y_m)\leq \log(\sum_{i=1}^r T_i).$$

\emph{Case (iii): arbitrary $p(x_1)$.} Since $\mathcal V_0$ is the class of transient states, we have
$$\lim_{m\rightarrow\infty} \mathbb{P}[X_m\in \mathcal V_0]=0.$$
Thus, for any $\delta>0$, one can find some $m_0$ such that $\mathbb{P}[X_{m_0}\in \mathcal V_0]<\delta$. Let $U$ be an indicator function that $X_{m_0}\in \mathcal V_0$. Then,
 \begin{align*}I(X_1;Y_m)&\leq I(X_{m_0}; Y_m)\\&=
I(UX_{m_0}; Y_m)
\\&\leq H(U)+I(X_{m_0}; Y_m|U)
\\&\leq h(\delta)+\delta I(X_{m_0}; Y_m|U=0)+(1-\delta) I(X_{m_0}; Y_m|U=1)
\\&\leq h(\delta)+\delta \log|\mathcal X|+(1-\delta) I(X_{m_0}; Y_m|U=1)
\end{align*}
Now, conditioned on $U=1$, the pmf of $p(x_{m_0}|u=1)$ falls in the class of pmfs studied in case (ii). Therefore, $\lim_{m\rightarrow\infty}  I(X_{m_0}; Y_m|U=1)\leq \log(\sum_{i=1}^r T_i)$. Hence,
 \begin{align*}\lim_{m\rightarrow\infty} I(X_1;Y_m)&\leq h(\delta)+\delta \log|\mathcal X|+(1-\delta)  \log(\sum_{i=1}^r T_i).
\end{align*}
We obtain the desired result by letting $\delta$ tend to zero.


\begin{thebibliography}{100}

\bibitem{Akyldiz1}
M. Pierobon, I. F. Akyildiz, ``A physical end-to-end model for molecular communication in nanonetworks," \emph{IEEE Journal on Selected Areas in Communications}, 28 (4): 602--611, 2010.


\bibitem{Nakano11}
T. Nakano, M. J.  Moore, F. Wei, A. V. Vasilakos, J. Shuai, ``Molecular communication and networking: Opportunities and challenges,"  \emph{IEEE Transactions on NanoBioscience}, 11(2): 135--148, 2012.



\bibitem{NarimanSurvey}
N. Farsad, H. B. Yilmaz, A. Eckford, C. B. Chae, and W. Guo, ``A Comprehensive survey of recent advancements in molecular communication," to appear in  \emph{IEEE Communications Surveys \& Tutorials}, arXiv:1410.4258v3, Feb 2016.


\bibitem{Gastpar}
M. Gastpar, B. Rimoldi, M. Vetterli, ``To code, or not to code: Lossy source-channel communication revisited," \emph{IEEE Transactions on Information Theory}, 49(5): 1147--1158, 2003.

\bibitem{Ourwork1}
R. Mosayebi, H. Arjmandi, A. Gohari, M. Nasiri-Kenari and U. Mitra, ``Receivers for diffusion-based molecular communication: Exploiting memory and sampling rate," \emph{IEEE Journal on Selected Areas in Communications}, 32 (12): 2368--2380, 2014.

\bibitem{Ourwork2}
M. Movahednasab, M. Soleimanifar, A. Gohari, M. Nasiri-Kenari and U. Mitra, ``Adaptive Transmission Rate With a Fixed Threshold Decoder for Diffusion-Based Molecular Communication," \emph{IEEE Transactions on Communications}, 64 (1): 236--248, 2015.


\bibitem{Yao82a}
A. C. Yao, ``Theory and applications of trapdoor functions,"  \emph{23rd Annual IEEE Symposium on Foundations of Computer Science}, 80-–91, 1982.


\bibitem{HILL99}
J. Hastad, R. Impagliazzo, L. A. Levin and M. Luby, ``A pseudorandom generator from any one-way function," \emph{SIAM Journal on Computing}, 28 (4): 1364--1396, 1999.


\bibitem{ElGamalGreene}
 A. El Gamal, J. Greene, and K. Pang, ``VLSI complexity of coding," the \emph{MIT Conf. on Adv. Research in VLSI}, Cambridge, MA, Jan. 1984.


\bibitem{Grover1}
P. Grover, A. Goldsmith and A. Sahai, ``Fundamental limits on the power consumption of encoding and decoding,"  \emph{Proc. IEEE Int. Symp. on Inf. Theory (ISIT)}, 2716--2720, 2012.

\bibitem{Grover2}
P. Grover, ``Information Friction and Its Implications on Minimum Energy Required for Communication,"  \emph{IEEE Transactions on Information Theory}, 61(2): 895--907, 2015.


\bibitem{EinSarFekISIT12}
A. Einolghozati, M. Sardari, and F. Fekri, ``Collective sensing-capacity of bacteria populations," \emph{Proc. IEEE Int. Symp. on Inf. Theory (ISIT)}, 2959--2963, 2012.



\bibitem{FarsadGoldsmith}
N. Farsad and A. Goldsmith, ``A molecular communication system using acids, bases and hydrogen ions," 2016 IEEE 17th International Workshop on In Signal Processing Advances in Wireless Communications (SPAWC), 1--6, 2016.





\bibitem{MosayebiIWCIT16}
R. Mosayebi, A. Gohari, M. Mirmohseni, and M. Nasiri-Kenari, ``Type based sign modulation for molecular communication," to appear in \emph{Proc. Iran Workshop on Communication and Information Theory (IWCIT)}, 2016.


\bibitem{FutureIT}
J.G. Andrews, A. Dimakis, L. Dolecek, M. Effros, M. Medard, O. Milenkovic, A. Montanari, S. Vishwanath, E. Yeh, R. Berry, K. Duffy, ``A Perspective on Future Research Directions in Information Theory," arXiv preprint arXiv:1507.05941, 2015.




\bibitem{Andrews}
B. W. Andrews and P. A. Iglesias, ``An information-theoretic characterization of the optimal gradient sensing response of cells," \emph{PLoS Computational Biology}, 3(8):e153, 2007.




\bibitem{PierobonAkyildiz2011}
M. Pierobon, I. F. Akyildiz, ``Diffusion-based noise analysis for molecular communication in nanonetworks," IEEE Transactions on Signal Processing, 59 (6), 2532-2547, 2011.


\bibitem{Chou2015}
C.T. Chou, ``A Markovian approach to the optimal demodulation of
diffusion-based molecular communication networks," IEEE Transactions
on Communications, vol. 63, no,10, pp. 3728-3743, Oct. 2015.



\bibitem{Biobook}
P. C. Bressloff,  \emph{Stochastic processes in cell biology}, Heidelberg: Springer, Switzerland, 2014.




\bibitem{ArjAhmSchKen16}
H.~Arjmandi, A.~Ahmadzadeh, R.~Schober, and M.~Nasiri-Kenari, ``Ion Channel Based Bio-Synthetic Modulator for Diffusive Molecular Communication,'' to appear in \emph{IEEE Transactions on NanoBioscience}, 2016.


\bibitem{Mahdavifar}
H. Mahdavifar, A. Beirami, ``Diffusion channel with Poisson reception process: capacity results and applications," \emph{Proc. IEEE Int. Symp. on Inf. Theory (ISIT)}, 1956--1960, 2015.





\bibitem{Pao}
C. V. Pao,  ``Nonlinear parabolic and elliptic equations," 1992, Plenum Press, New York

\bibitem{Oksendal}
B. Oksendal, ``Stochastic differential equations: an introduction with applications," Springer-Verlag Berlin Heidelberg, 2003.


\bibitem{Feller}
W. Feller, ``An introduction to probability theory and its applications: volume II," London-New York-Sydney-Toronto: John Wiley \& Sons, 1971.


\bibitem{Protter}
P. E. Protter, ``Stochastic Differential Equations. In Stochastic Integration and Differential Equations," Springer Berlin Heidelberg, 2005.



\bibitem{Roisin}
B. Cushman-Roisin, ``Environmental transport and fate," Thayer School of Engineering Dartmouth College, University Lecture, 2012.





\bibitem{Arjmandi2013}
H.~Arjmandi, A.~Gohari, M.~Nasiri-Kenari, and F.~Bateni, ``Diffusion-based nanonetworking: A new modulation technique and performance analysis,'' \emph{Communications Letters, IEEE}, 17 (4): 645--648, 2013.

\bibitem{PieAky11}
M. Pierobon and I. Akyildiz, ``Diffusion-based noise analysis for molecular communication in nanonetworks,'' \emph{IEEE Transactions on Signal Processing}, 59 (6):  2532--2547, 2011.

\bibitem{MahMakMou14}
M. Mahfuz, D. Makrakis, and H. Mouftah, ``A comprehensive study of sampling-based optimum signal detection in concentration-encoded molecular communication,'' \emph{IEEE Transactions on NanoBioscience,} 13 (3):  208--222, 2014.

\bibitem{NoeCheSch14}
A. Noel, K. Cheung, and R. Schober, ``Improving receiver performance of diffusive molecular communication with enzymes,'' \emph{IEEE Transactions on NanoBioscience}, 13 (1): 31--43, 2014.

\bibitem{YilHerTugCha14}
H. Yilmaz, A. Heren, T. Tugcu, and C.-B. Chae, ``Three-dimensional channel characteristics for molecular communications with an absorbing receiver,'' \emph{IEEE Communication Letters}, 18 (6): 929-932, 2014.

\bibitem{Ghavami}
S. Ghavami, R. S. Adve, and F. Lahouti, ``Information rates of ask-based molecular communication in fluid media," \emph{IEEE Transactions on Molecular, Biological and Multi-Scale Communications}, 1 (3): 277–291, 2015.



\bibitem{RossBook}
S. M. Ross,  \emph{Introduction to pobability models (tenth edition),} Academic Press, 2010.



\bibitem{Timing4n}
N. Farsad, Y. Murin, A. Eckford, A. Goldsmith, ``Capacity limits of diffusion-based molecular timing channels," arXiv preprint arXiv:1602.07757, 2016.


\bibitem{Timing3}
K. V. Srinivas, A. W. Eckford, R. S. Adve, ``Molecular communication in fluid media: The additive inverse gaussian noise channel,"  \emph{IEEE Transactions on Information Theory}, 58(7), 4678--4692, 2012.


\bibitem{TimingAux1}
H. B. Yilmaz, A. C. Heren, T. Tugcu, and C.-B. Chae, ``Three- dimensional channel characteristics for molecular communications with an absorbing receiver," \emph{IEEE Commuminications Letters}, 18(6): 929–-932, 2014.


\bibitem{AkanNew}
D. Kilinc, O. B. Akan, ``Receiver design for molecular communication," \emph{IEEE Journal on  Selected Areas in Communications}, 31 (12): 705-714, 2013.


\bibitem{AkanReciever}
B. Atakan, O. B. Akan, ``An information theoretical approach for molecular communication," \emph{Bio-Inspired Models of Network, Information and Computing Systems, Bionetics}  33--40, 2007.

\bibitem{AtaAka09}
B. Atakan and O. B. Akan, ``Single and multiple-access channel capacity in molecular nanonetworks," \emph{Nano-Net}, ser. Springer Lect. Notes Inst. Comput. Sci., Social Inf. and Telecommu. Eng., A. Schmid, S. Goel, W. Wang, V. Beiu, and S. Carrara, Eds., 20: 14--23, 2009.


\bibitem{EinSarFekITW11}
A. Einolghozati, M. Sardari, and F. Fekri, ``Capacity of diffusion-based molecular communication with ligand receptors," \emph{Proc. IEEE Inf. Theory Workshop (ITW)}, 85--89, 2011.



\bibitem{Ligand1}
A. W. Eckford,  P. J. Thomas,  ``Capacity of a simple intercellular signal transduction channel",
\emph{Proc. IEEE Int. Symp. on Inf. Theory (ISIT)}, 1834--1838, 2013.


\bibitem{Ligand2}
M. Tahmasbi, F. Fekri, ``On the capacity achieving probability measures for molecular receivers",  \emph{Proc. IEEE Inf. Theory Workshop (ITW)}, 109--113, 2015.



\bibitem{ReactiveReceiver}
A. Ahmadzadeh, H. Arjmandi, A. Burkovski, R. Schober, ``Comprehensive Reactive Receiver Modeling for Diffusive Molecular Communication Systems: Reversible Binding, Molecule Degradation, and Finite Number of Receptors," IEEE Transactions on NanoBioscience, 2016.



\bibitem{PieAky11s}
M. Pierobon and I. Akyildiz, ``Noise analysis in ligand-binding reception for molecular communication in nanonetworks,'' \emph{IEEE Transactions on Signal Processing}, 59(9): 4168--4182,  2011.

\bibitem{Cho15}
C. T. Chou, ``Impact of receiver reaction mechanisms on the performance of molecular communication networks,'' \emph{IEEE Transactions on Nanotechnology}, 14 (2): 304--317,  2015.


\bibitem{Heren}
A. C. Heren, H. B. Yilmaz, C. B. Chae, T. Tugcu, ``Effect of Degradation in Molecular Communication: Impairment or Enhancement?," \emph{IEEE Transactions on Molecular, Biological and Multi-Scale Communications},  1(2): 217--229, 2015.


\bibitem{Timing1}
A. W. Eckford, ``Nanoscale communication with brownian motion," in \emph{Proc. Conf. on Inf. Sci. and Syst. (CISS)}, 160-–165, 2007.



\bibitem{Timing2/52}
C. Rose, C., I. S. Mian, ``Signaling with identical tokens: Lower bounds with energy constraints,"  \emph{Proc. IEEE Int. Symp. on Inf. Theory (ISIT)}, 1839--1843, 2013.


\bibitem{Timing6}
C. Rose and I. Mian, ``A fundamental framework for molecular communication channels: Timing \& payload," \emph{Proc. IEEE Int. Conf. on Commun. (ICC)}, 1043--1048, 2015.



\bibitem{Arifler}
D. Arifler, ``Capacity analysis of a diffusion-based short-range molecular nano-communication channel," \emph{Computer Networks}, 55(6), 1426-1434, 2011.


\bibitem{AmiArjGohNasMit15}
G.~Aminian, H.~Arjmandi, A.~Gohari, M.~Nasiri-Kenari, and U.~Mitra, ``Capacity of diffusion based molecular communication networks in the {LTI-Poisson} model,''{\em IEEE Transactions on Molecular, Biological and Multi-Scale Communications,} 1(2): 188--201, 2015.


\bibitem{Gallager}
R. G. Gallager, Information  Theory and Reliable Communication. New York Wiley, 1968.


\bibitem{Massey}
W.~Hirt and J.~L. Massey, ``Capacity of the discrete-time {Gaussian} channel
  with intersymbol interference,''  \emph{IEEE Transactions on Information Theory},  34(3):38--38, 1988.

\bibitem{Shamai}
S. Shamai, L. Ozarow, and A. Wyner, ``Information rates for a discretetime
gaussian channel with intersymbol interference and stationary
inputs,”   \emph{IEEE Transactions on Information Theory}, 37(6): 1527–1539, 1991.



\bibitem{AtakanAKan08}
B. Atakan and O. B. Akan, ``On channel capacity and error compensation in molecular communication," \emph{Trans. Comput. Syst. Biol. X}, ser. Springer Lect.
Notes Comput. Sci., C. Priami, F. Dressler, O. B. Akan, and A. Ngom, Eds., 5410: 59--80, 2008.



\bibitem{LiuYan14}
Q. Liu, and K. Yang, ``Channel capacity analysis of a diffusion-based molecular communication system with ligand receptors," \emph{International Journal of Communication Systems}, 28(8):1508--1520, 2014.


\bibitem{EinSarFekTWC13}
A. Einolghozati, M. Sardari, and F. Fekri, ``Design and analysis of wireless communication systems using diffusion-based molecular communication among bacteria," \emph{IEEE
Transactions on Wireless Communications}, 12(12): 6096--6105, 2013.

\bibitem{EinSarFekISIT13}
A. Einolghozati, M. Sardari, and F. Fekri, ``Relaying in diffusion-based molecular communication," \emph{Proc. IEEE Int. Symp. on Inf. Theory (ISIT)}, 1844--1848, 2013.

\bibitem{EinSarFekICC14}
A. Einolghozati, M. Sardari, and F. Fekri, ``Decode and forward relaying in diffusion-based molecular communication between two populations of biological agents," \emph{Proc. IEEE Int. Conf. on Commun. (ICC)}, 3975--3980, 2014.



\bibitem{Kuran10}
M. S. Kuran, H. B. Yilmaz, T. Tugcu, and B. Ozerman, ``Energy model for communication via diffusion in nanonetworks," \emph{Elsevier Nano Commun. Netw.}, 1(2): 86--95,  2010.



\bibitem{Lapidoth2011}
A.~Lapidoth, J.~H. Shapiro, V.~Venkatesan, and L.~Wang, ``The discrete-time
  poisson channel at low input powers,'' \emph{IEEE Transactions on Information Theory}, 57(6): 3260--3272, 2011.


\bibitem{LapidothMoser2009}
A.~Lapidoth and S.~M. Moser, ``On the capacity of the discrete-time poisson
  channel,''  \emph{IEEE Transactions on Information Theory}, 55(1):303--322, 2009.




\bibitem{AkyldizPoisson}
Y. Chahibi and I. F. Akyildiz, ``Molecular Communication Noise and Capacity Analysis for Particulate Drug Delivery Systems", \emph{IEEE Transactions on Communications}, 62 (11), 3891--3903, 2014.



  \bibitem{Verdu}
S.~Verdu, ``Multiple-access channels with memory with and without frame
  synchronism,''  \emph{IEEE Transactions on Information Theory}, 35(3): 605--619, 1989.



\bibitem{VerduLautum}
D. P. Palomar, S. Verdu,  ``Lautum information,"  \emph{IEEE Transactions on Information Theory}, 54(3): 964--975, 2008.




\bibitem{Mitzenmacher}
M. Mitzenmacher and E. Upfal, \emph{Probability and computing: randomized algorithms and probabilistic analysis}, Cambridge University Press, 2005. 





\bibitem{EinSarBeiFekISIT11}
A. Einolghozati, M. Sardari, A. Beirami, and F. Fekri, ``Capacity of discrete molecular diffusion channels," \emph{Proc. IEEE Int. Symp. on Inf. Theory (ISIT)}, 723--727, 2011.


\bibitem{Venkat}
V. Anantharam and S. Verdu, ``Bits through queues,”  \emph{IEEE Transactions on Information Theory},  42(1): 4-–18, 1996.


\bibitem{Timing2/5}
C. Rose, C., I. S. Mian, ``Signaling with identical tokens: Upper bounds with energy constraints,"  \emph{Proc. IEEE Int. Symp. on Inf. Theory (ISIT)},  1817--1821, 2014.

\bibitem{Timing2}
L. Cui, A. W. Eckford, ``The delay selector channel: Definition and capacity bounds," \emph{IEEE Canadian Workshop on  Information Theory (CWIT)},  15--18, 2011.

\bibitem{Topsoe}
F. Topsoe, ``An information theoretical identity and a problem involving capacity," \emph{Studia Scientiarum Math. Hungarica}, 2:291–292, 1967.


\bibitem{Timing4}
M. N. Khormuji, ``On the capacity of molecular communication over the AIGN channel,"  \emph{Proc. Conf. on Inf. Sci. and Syst. (CISS)}, 1--4, 2011.

\bibitem{Timing5}
H. Li, S. Moser, and D. Guo, ``Capacity of the memoryless additive inverse gaussian noise channel," \emph{IEEE Journal on Selected Areas in Communications}, 32(12): 2315--2329,  2014.

\bibitem{Timing5n}
W. Schwarz, ``On the convolution of inverse Gaussian and exponential random variables," \emph{Communications in Statistics — Theory and Methods},  31(12): 2113--2121, 2002.

\bibitem{EckfordChae}
A.W. Eckford, C. B. Chae ``Scaling laws for molecular communication",   \emph{Proc. IEEE Int. Symp. on Inf. Theory (ISIT)}, 1281--1285, 2014.

\bibitem{Timing10}
M. Kovacevic, P.  Popovski, ``Zero-error capacity of a class of timing channels,"  \emph{IEEE Transactions on Information Theory}, 60(11), 6796--6800, 2014.





\bibitem{VerduHan}
S. Verdu, T. S. Han, ``A general formula for channel capacity,"  \emph{IEEE Transactions on Information Theory}, 40 (4), 1147-1157,  1994.















\bibitem{Xie}
Q. Xie and A. R. Barron, ``Minimax redundancy for the class of memoryless
sources,"  \emph{IEEE Transactions on Information Theory}, 43(2): 646--657, 1997.


\bibitem{AmiFarMirNasFek16}
G.~Aminian, M. Farahnak-Ghazani, M. Mirmohseni, M.~Nasiri-Kenari, and F. Fekri, ``On the Capacity of Point-to-Point and Multiple-Access Molecular Communications with Ligand-Receptors,'' to appear in {\em IEEE Transactions on Molecular, Biological and Multi-Scale Communications,} 2016.



\bibitem{ThomasEckford}
P. J. Thomas, A. W. Eckford, A. W. ``Shannon Capacity of Signal Transduction for Multiple Independent Receptors," arXiv preprint arXiv:1604.03508, 2016.

\bibitem{Chen05}
J. Chen and T. Berger, “The capacity of finite-state Markov channels with feedback,”  \emph{IEEE Transactions on Information Theory},  51(3): 780--798, 2005.

\bibitem{Permuter14}
H. H. Permuter, H. Asnani, and T. Weissman, ``Capacity of a POST channel with and without feedback,”  \emph{IEEE Transactions on Information Theory}, 60(10): 6041--6057, 2014.



\bibitem{EckfordThomasNew}
P. J. Thomas, A. Eckford ``Shannon Capacity of Signal Transduction for Multiple Independent Receptors," IEEE International Symposium on Information Theory (ISIT), 1804--1808, 2016.



















\bibitem{Ubli}
N. Michelusi, S. Pirbadian, M.Y. El-Naggar, U. Mitra, ``A Stochastic Model for Electron Transfervin Bacterial Cables", \emph{IEEE Journal on Selected Areas in Communication}, 32(12): 2402--2416, 2014.


\bibitem{Kramer}
G. Kramer, M. I. Yousefi and F. R. Kschischang, ``Upper bound on the capacity of a cascade of nonlinear and noisy channels," \emph{IEEE Information Theory Workshop (ITW)}, 1-4, 2015.




\bibitem{cas__ref1}
R. B. Ash, \emph{Information Theory}, New York Wiley, 1965

\bibitem{cas__ref2}
T. M. Cover and J. A. Thomas, \emph{Elements of Information Theory,} New York Wdey, 1991.

\bibitem{cas__ref3}
R. J. McEliece, \emph{The Theory of Information and Coding,} Reading, MA: Addison-Wesley, 1977

\bibitem{cascade_1}
Aaron B. Kiely and John T. Coffey, ``On the capacity of a cascade of channels", \emph{IEEE Transactions on Information Theory}, 39 (4):1310--1321, 1993.

\bibitem{Majani}
E. Majani, ``A model for the study of very noisy channels, and applications," \emph{Ph.D. dissertation}, California Inst. of Technol., Pasadena,
CA, 1988.

\bibitem{Simon}
M.K. Simon, ``On the capacity  of a cascade  of  identical discrete memoryless nonsingular channels,"   \emph{IEEE Transactions on Information Theory}, 16(1): 100-102, 1970.

\bibitem{Silverman}
R.A. Siveman, ``On binary channels and their cascades," \emph{IRE Transactions on Information Theory}, 1(3): 19-27, 1955.


\bibitem{AG}
R. Ahlswede and P. Gacs, ``Spreading of sets in product spaces and hypercontraction of
the Markov operator," \emph{Annals of Probability},  925–-939, 1976.

\bibitem{AGKN13}
V. Anantharam, A. Gohari, S. Kamath, and C. Nair. ``On maximal
correlation, hypercontractivity, and the data processing inequality studied by Erkip
and Cover," arXiv preprint arXiv:1304.6133, 2013.

\bibitem{Polyanskiy}
Y. Polyanskiy and Y. Wu, ``Dissipation of information in channels with input constraints,"   \emph{IEEE Transactions on Information Theory}, 62 (1), 35--55, 2016.

\bibitem{Line}
U. Niesen, C. Fragouli, D. Tuninetti, ``On the capacity of line networks,"  \emph{IEEE Transactions on Information Theory}, 53 (11), 4039--4058, 2007.




\bibitem{svl13}
R. Subramanian, B. N. Vellambi, and I. Land. ``An improved bound on information loss
due to finite block length in a Gaussian line network,"
 \emph{Proc. IEEE Int. Symp. on Inf. Theory (ISIT)}, 1864--1868, 2013.

\bibitem{sub12}
 R. Subramanian, ``The relation between block length and reliability for a cascade of
AWGN links," \emph{Proc. 2012 Int. Zurich Seminar on Communications (IZS)},
71–-74, 2012.


\bibitem{energyLu}
Y. Lu, M. D. Higgins, M. S. Leeson, ``Comparison of channel coding schemes for Molecular Communications Systems," \emph{IEEE Transactions on Communications},  63 (11): 3991 -- 4001, 2015.


\bibitem{EnergyModelFarsad}
N. Farsad, H. B. Yilmaz, C. B. Chae, A. Goldsmith, ``Energy Model for Vesicle-Based Active Transport Molecular Communication,"  arXiv:1510.05075 (2015).


\bibitem{Media-Comm}
A. K. Khandani, ``Media-based modulation: A new approach to wireless transmission,"
\emph{Proc. IEEE Int. Symp. on Inf. Theory (ISIT)}, 3050--3054, 2013.


\bibitem{NakanoSuda}
T. Nakano, T. Suda, M. J. Moore, ``Molecular Communication through Biological Pattern Formation," \emph{IEEE Global Communications Conference (GLOBECOM)}, 1--7, 2015.


\bibitem{Fibo}
C. A. Charalambides, \emph{Enumerative combinatorics,} CRC Press, 2002.














\end{thebibliography}
\end{document}